 \documentclass[12pt,preprint]{aastex}
\usepackage{natbib}
\usepackage{amsmath}
\usepackage{hyperref}
\usepackage{graphicx}
\usepackage{mathtools}
\usepackage{rotating}
\usepackage{placeins}

 \usepackage{lineno}

\newcommand{\ltsima} {$\; \buildrel < \over \sim \;$}
\newcommand{\gtsima} {$\; \buildrel > \over \sim \;$}
\newcommand{\lta} {\lower.5ex\hbox{\ltsima}}
\newcommand{\gta} {\lower.5ex\hbox{\gtsima}}

\slugcomment{Date: \today}

\begin{document}

\title{Localization of Gamma-Ray Bursts using the {\it Fermi} Gamma-Ray Burst Monitor}
\shorttitle{Fermi GBM GRB Localizations}
\shortauthors{Connaughton et al.}

\author{V. Connaughton\altaffilmark{*,1}, M. S. Briggs\altaffilmark{1}, A. Goldstein\altaffilmark{2},
C.~A.~Meegan\altaffilmark{3},
W.~S.~Paciesas\altaffilmark{4},
R.~D.~Preece\altaffilmark{5},
C.~A.~Wilson-Hodge\altaffilmark{2},
M.~H.~Gibby\altaffilmark{6},
J.~Greiner\altaffilmark{7},
D.~Gruber\altaffilmark{8},
P.~Jenke\altaffilmark{3},
R.~M.~Kippen\altaffilmark{9}.
V.~Pelassa~\altaffilmark{3},
S.~Xiong\altaffilmark{3},
H.-F.~Yu\altaffilmark{7,10},
P.~N.~Bhat\altaffilmark{3},
J.~M.~Burgess\altaffilmark{1},
D.~Byrne\altaffilmark{11},
G.~Fitzpatrick\altaffilmark{11},
S.~Foley\altaffilmark{11},
M.~M.~Giles\altaffilmark{6},
S.~Guiriec\altaffilmark{12},
A.~J.~van~der~Horst\altaffilmark{13},
A.~von~Kienlin\altaffilmark{7},
S.~McBreen\altaffilmark{11},
S.~McGlynn\altaffilmark{11},
D.~Tierney\altaffilmark{11}, \&
B.-B.~Zhang\altaffilmark{3}}

\altaffiltext{*}{Email: valerie@nasa.gov}
\altaffiltext{1}{CSPAR and Physics Dept, University of Alabama in Huntsville, 320 Sparkman Dr., Huntsville, AL 35899, USA}
\altaffiltext{2}{Astrophysics Office, ZP12, NASA/Marshall Space Flight Center, Huntsville, AL 35812, USA}
\altaffiltext{3}{CSPAR, University of Alabama in Huntsville, 320 Sparkman Dr., Huntsville, AL 35899, USA}
\altaffiltext{4}{Universities Space Research Association, Huntsville, AL, USA}
\altaffiltext{5}{Dept. of Space Science, University of Alabama in Huntsville, 320 Sparkman Dr., Huntsville, AL 35899, USA}
\altaffiltext{6}{Jacobs Technology, Inc., Huntsville, AL, USA}
\altaffiltext{7}{Max-Planck-Institut f\"ur extraterrestrische Physik, Giessenbachstrasse 1, 85748 Garching, Germany}
\altaffiltext{8}{Planetarium S\"udtirol, Gummer 5, 39053 Karneid, Italy}
\altaffiltext{9}{Los Alamos National Laboratory, NM 87545, USA}
\altaffiltext{10}{Excellence Cluster Universe, Technische Universit\"at M\"unchen, 
  Boltzmannstr. 2, 85748, Garching, Germany}
\altaffiltext{11}{School of Physics, University College Dublin, Belfield, Stillorgan Road, Dublin 4, Ireland}
\altaffiltext{12}{NASA Goddard Space Flight Center, Greenbelt, MD 20771, USA}
\altaffiltext{13}{Astronomical Institute, University of Amsterdam, Science Park 904, 1098 XH Amsterdam, The Netherlands}

\begin{abstract}
The {\it Fermi} Gamma-ray Burst Monitor (GBM) has detected over 1400 Gamma-Ray Bursts (GRBs) since it began science operations in July, 2008.
We use a subset of over 300 GRBs localized by 
instruments such as {\it Swift}, the {\it Fermi} Large Area Telescope, INTEGRAL, and MAXI, or through triangulations
from the InterPlanetary Network (IPN), to analyze the accuracy of GBM GRB localizations.
We find that the reported statistical uncertainties on GBM localizations, which 
can be as small as $1^\circ$, underestimate the distance of the GBM positions to
the true GRB locations and we attribute this to systematic uncertainties.  The
distribution of systematic uncertainties is well represented 
(68\% confidence level) by a $3.7^\circ$ Gaussian with
a non-Gaussian tail that contains about 10\% of GBM-detected GRBs and extends to approximately $14^\circ$.
A more complex model suggests that there is a dependence of the systematic uncertainty 
on the position of the GRB in spacecraft coordinates, with GRBs in the quadrants on the Y-axis better localized
than those on the X-axis.
\end{abstract}


\section{Introduction}
In four years of operation, the {\it Fermi} Gamma-ray Space Telescope has opened a new window to the world of Gamma-Ray Burst (GRB) spectroscopy.
Observations by the Gamma-Ray Burst Monitor (GBM, \cite{meegan09}) between 8 keV and 40 MeV and the Large Area Telescope (LAT, \cite{atwood09})
from 20 MeV to hundreds of GeV have provided a high-energy view over an unprecedentedly broad energy baseline. 
Follow-up observations of LAT-detected GRBs have revealed the redshift of about a dozen
 GRBs detected above 100 MeV (\cite{lat_catalog} and references therein; \cite{gcn14685, gcn14455, gcn14983, gcn15187}).  
Determining the redshift of GRBs enables the study of the energetics
and rest-frame properties of these events.
Owing to the localization limitations of the GBM experiment, follow-up observations of 
GRBs localized at trigger time only by GBM were rare until we disseminated the
results of the work presented here, in which 
uncertainties in GRB localizations are characterized.
GRB~090902B \citep{apj090902b} was observed by ROTSE \citep{pandey090902b}
an hour after the trigger, with ROTSE tiling the GBM error circle 
hours before the burst was better localized using LAT and {\it Swift} data. 
A source was subsequently found in the ROTSE data at the position of the burst, providing
the earliest measurements of the GRB afterglow.  
GRB~130702A was observed by the intermediate Palomar Transient Factory (iPTF), with the
telescope tiling 72 square degrees in 10 separate pointings, and uncovering 
the afterglow of the GRB 4.2 hours after the GBM trigger \citep{singer2013}.  This was the first 
discovery of afterglow emission from observations that used only the GBM localization and did not
require additional, more accurate, positions from the LAT or from {\it Swift} to find the source in
the large observed sky region.
Since then, regular observations 
by iPTF of GRBs localized by GBM revealed the afterglows for eight more GRBs 
in 35 attempts (L. Singer et al., 
in preparation). 
 In general, however, the small fields-of-view of the most
sensitive follow-up telescopes have deterred regular observations of the degrees-scale uncertainty regions
resulting from GBM localizations.  A further discouraging aspect of GBM localizations is that
the total error is often larger than the reported statistical uncertainty. 
Using a Bayesian method similar to that reported here, \citet{briggs2008} analyzed 36 GBM GRB localizations and
found the 68\% confidence level systematic uncertainty to be $3.8 \pm 0.5^\circ$.  \citet{ipn_catalog} 
use a sample of 149 GRBs detected by GBM and by other instruments in the InterPlanetary Network (IPN) to
infer a 90\% confidence level systematic uncertainty of $6^\circ$.  
We report here a study of systematic uncertainties using over 300 reference locations provided by
other instruments and by the IPN.   The technique we use is based on the
work of \cite{carlo96} and \cite{briggs99}, who developed a Bayesian approach to characterize
systematic uncertainties for the GRBs detected by the Burst And Transient Source Experiment (BATSE) on the
Compton Gamma-Ray Observatory.

GBM detects about 240 GRBs per year \citep{paciesas_1grb, azk_2grb}, providing 
real-time locations for the follow-up community.
Approximately 17\% of {\it Fermi} GBM-detected
GRBs are short in duration \citep{azk_2grb}.  Unlike the 
short GRB population detected by {\it Swift}, which may contain weak 
collapsar events \citep{bromberg13}, it is likely 
that most GBM-detected short GRBs (SGRBs) are associated with the merger of compact binary systems. There are therefore
40 or so merger GRBs per year detected by GBM that could potentially be observed at other wavelengths.  Mergers of compact binary
systems are likely sources of gravitational waves (GW).  No associations were found between 154 GRBs detected in
2009-2010 by various spacecraft,
including GBM, and potential signals in the LIGO and Virgo GW experiments \citep{abadie12}. This was not unexpected given the
detection horizon of about 20 Mpc for these experiments. Advanced configurations of both experiments will be deployed over
the next few years, with horizons of 400 [1000] Mpc for NS-NS [NS-BH] mergers \citep{abadie10}.  It is realistic to expect
several joint detections of SGRBs by GBM and GW candidates by A-LIGO/Virgo per year \citep{vc_multimessenger}. For
the first years of A-LIGO/Virgo operation, localization
uncertainties for GW candidates are estimated to be 1000 square degrees or more \citep{abadie2013},
larger than the GBM uncertainty regions, consisting of annuli segments that may encompass non-contiguous 
sky regions.  It is especially important to enable the optical community to observe SGRBs 
while the afterglow is still bright enough to be detected above the background in a large error box. 
For GBM-detected SGRBs, this may imply covering only a small part of the error box.  

In Section 2 we describe the detection and localization of GRBs by GBM and in Section 3 their
dissemination to the public over the GRB Coordinates Network.  The sample of reference locations
from detections by other instruments or by the InterPlanetary Network is introduced in Section 4, 
where reference point locations are compared to the GBM localizations and their reported
statistical uncertainties. In section 5, we describe our method to assess the effect of
systematic uncertainties on GRB localization. We report results for the models we tested. In section
6 we summarize our results and describe new data products that use these results to facilitate
follow-up observations of GRBs localized by GBM.

\section{Localization of GRBs by GBM}
\begin{figure}
\center{
 \includegraphics[width=3in]{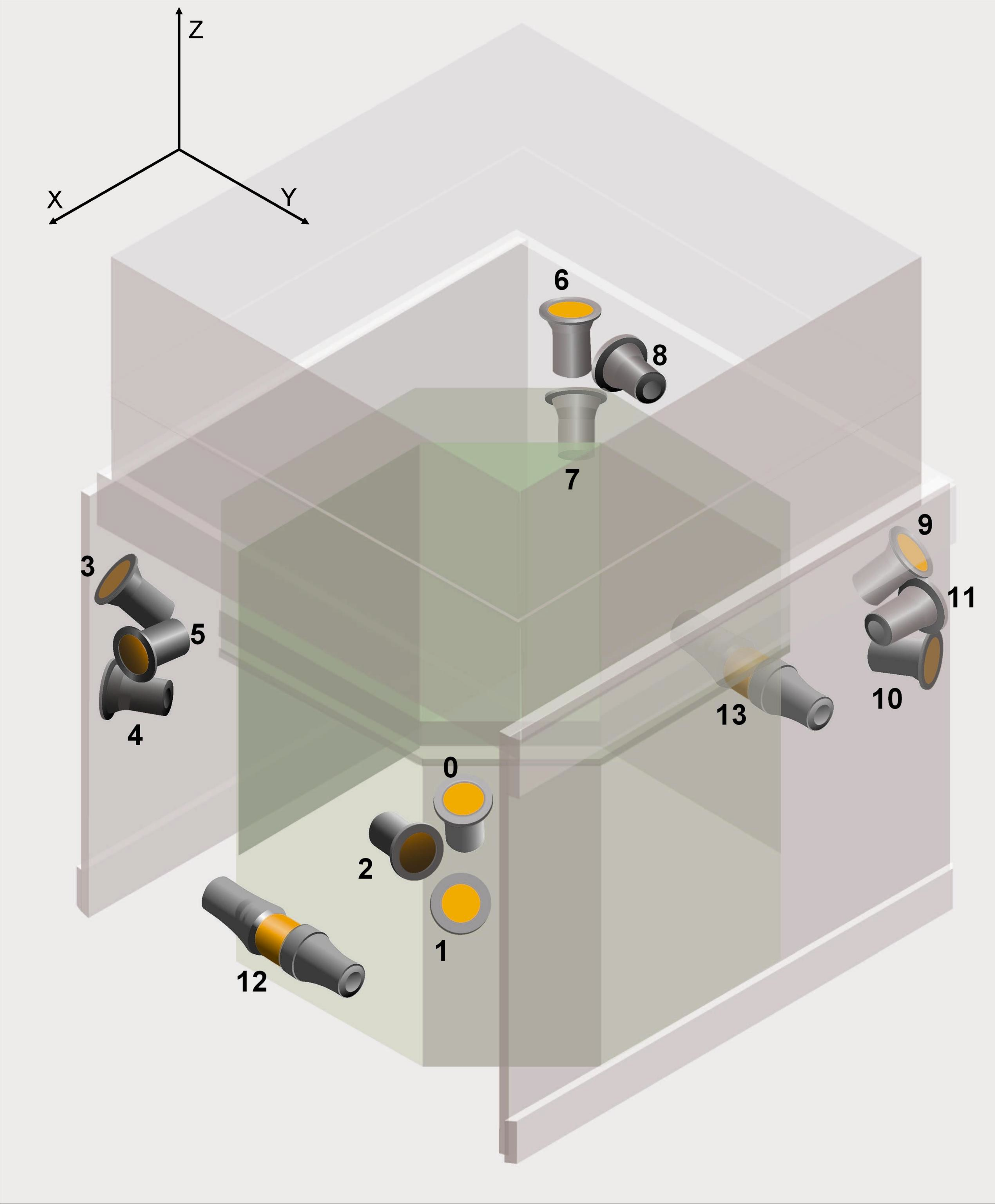}
 \caption{
In the spacecraft coordinate system, the Z-axis is aligned with the LAT pointing.  
The 14 NaI detectors and 2 BGO detectors are mounted on the +ve and -ve X-axes. The NaI detectors are numbered
0-11 and the BGO detectors 0 and 1. The solar panels and LAT radiators are mounted on the +ve and -ve Y-axes.   
\label{fig:det_place}}}
 \end{figure}

The GBM views the entire unocculted sky, over 7 steradians, using 12 sodium iodide (NaI) detectors, 
1.27~cm thick and 12.7~cm in diameter,
covering an energy range from 8 keV to 1 MeV, and two bismuth
germanate (BGO) scintillators, 12.7~cm in diameter and thickness, placed on opposite sides of the
spacecraft, with energy coverage from 200 keV to 40 MeV \citep{meegan09}. 
In a coordinate system centered on the spacecraft, the
Z-axis is oriented along the boresight of the LAT, the X-axis joins the two BGO detectors,
and the Y-axis joins the LAT radiators, as shown in Figure~\ref{fig:det_place}.  
The placement of the NaI detectors, which all have different orientations, in four clusters of three detectors 
gives maximum coverage along the positive Z-axis.  This means GRBs in the LAT field-of-view are seen by
more GBM detectors than those outside.  If {\it Fermi} were pointed at the local zenith, the Earth would occupy the 
region in spacecraft coordinates viewed by the fewest detectors, along the negative Z-axis.  In nominal sky-survey
mode, {\it Fermi} views the whole sky every 3 hours (2 spacecraft orbits) by tilting alternately north and south of the 
zenith each orbit to obtain uniform sky survey coverage when the exposure is averaged over just a few orbits.  
The angle of the tilt, called the rocking angle, has changed from $35^\circ$ at the start of the
{\it Fermi} mission to $50^\circ$ in October 2009, the change being necessary to place {\it Fermi} in a rocking profile that
keeps the spacecraft battery cool.

The energy information from the GBM detectors is binned in
128 channels constructed using the 4096
channels in the Data Processing Unit that result from the digitzation of the analog output from the detector front-end electronics.  
Channel-to-energy conversion uses pre-launch exposures
to radioactive sources between 14 keV and 4.4 MeV and a spline fit between and beyond the energies of the known sources 
covers the whole energy range.  Details of the energy calibration are given in \cite{bissaldi2009}.  
The localization of sources is done
with an 8-channel rebinning of the full-resolution 128-channel data into a quicklook data type that is downloaded in real-time when a trigger occurs.

Source localization uses the relative rates recorded in the 12 NaI detectors to
estimate the most likely arrival direction given the angular and spectral
response of the detectors.  
The detector response model was constructed from simulations 
incorporating the {\it Fermi} spacecraft mass model into GEANT4 \citep{rmk2007}. Incident photons with energies between
5~keV and $\sim$~50~MeV were injected from 272 directions in the spacecraft coordinate system to evaluate the geometry-dependent detector response.

A $\chi^2$ minimization process finds the direction on the sky from which the expected
count rates from our detector response model most closely match the observed detector count rates in all 12 NaI detectors.
The observed counts include a background component that is subtracted before modeling the source.
The observed source counts  are compared to expectation from evenly distributed
points on a 1$^\circ$ resolution grid in spacecraft coordinates that 
linearly
interpolates among the three closest of the native 272 sky directions 
giving 41168 grid points.  
The source rates contain three elements: direct flux from the source, flux scattered in
the spacecraft, and flux scattered from the atmosphere.  
Contributions from each of these components depend on the
observing geometry and on the spectrum of gamma rays from the source.  
We construct our instrument response matrices for
a particular GRB by adding two model terms, the direct response, which depends only on the source-spacecraft geometry, and 
the atmospheric response, which depends on the source-spacecraft-Earth geometry.  
The direct response is calculated by interpolating among the three closest  
of 272 sky points in the native database compiled from the GEANT simulations. 
Our atmospheric response calculation is a
simplification of the true geometry in which the Earth-spacecraft geometry has many solutions.   
We use the atmospheric response database established for BATSE \citep{locburst}. 
For each of the 
41168 grid points we calculate the rates normal to each detector
for the Earth-spacecraft geometry at the time of a trigger.  

We construct three tables with the model count rates between 50 and 300 keV for each detector
at each point on this grid, using three source spectra representing spectrally soft, medium, and hard
GRBs.  The source spectra are modeled using the Band function parametrization \citep{band}, 
two power-law components, $\alpha$ and $\beta$,
 that are smoothly joined, and a peak in the power per decade of energy, $E_{peak}$:


\[ F(E) = \left\{ 
\begin{array}{l l}
\left(\frac{E}{100}\right)^\alpha \exp\left[-\frac{(\alpha + 2)E}{E_{peak}}\right] \qquad  & E < [(\alpha - \beta) E_{peak}]/(\alpha + 2) \\ 
    = \left(\frac{E}{100}\right)^\beta \exp(\beta - \alpha)\left[\frac{(\alpha - \beta) E_{peak}}{100 (\alpha +2)}\right]^{\alpha - \beta} \qquad & 
 E \geq [(\alpha - \beta) E_{peak}]/(\alpha + 2)
\end{array}
\right.\]

The Band function parameters for these three spectra are $\alpha$, $\beta$, $E_{peak}$ = 
(-1.9, -3.7, 70 keV),  (-1, -2.3, 230 keV), and (0, -1.5, 1 MeV).  
Thus, the direct response is fixed and depends only on the source position in spacecraft coordinates. To this we add
an atmospheric component that is calculated during the execution of the localization code and that depends
 on the position of the Earth in spacecraft coordinates. 
A $\chi ^2$ minimization of each of the three tables (soft, medium, hard) relative to the observed rates produces the
most likely arrival direction in spacecraft coordinates for each of the three model spectra, 
and the lowest $\chi^2$ among the three minima is assumed to be from the spectrum that most closely resembles the burst.
The position from the selected table is translated to equatorial coordinates using the spacecraft attitude at trigger time.
The reported 68\% statistical uncertainty is the average distance to the grid points that lie at $\Delta \chi ^2$ = 2.3, assuming a circular
uncertainty region. A lower limit to the reported statistical uncertainty of $1^\circ$ is imposed to match the grid
 resolution, though in practice the $\chi^2$
contours can be very steep within the grid points.   
A discussion of the $\chi^2$ distributions and the selection of the best model based on $\chi^2$
is given in Appendix A.

\section{GBM Localization types}

It is desirable to generate GRB localizations as soon as possible after the trigger.
An automated process requiring no human intervention produces initial localizations 
from the on-board Flight Software algorithms (FSW locations) and on the ground (Ground-Auto locations), both within 10-30~s of the GRB trigger. 
Refined ground locations use more data and human judgment 
(Human in the Loop, or HitL locations) and are distributed on the order of an hour later.

When a GBM trigger occurs on-board {\it Fermi}, the GBM flight software (FSW) produces trigger data types that are downlinked to the ground
upon summoning the Tracking and Data Relay Satellite System (TDRSS) 
link, a 5 second process.  In addition to activating a TDRSS link, the GBM FSW communicates with the LAT
via the spacecraft bus, informing the LAT FSW that a trigger has occurred, its nature (GRB, solar flare etc.), and sends a localization produced
by the FSW.   Because of memory limitations on-board, the FSW localization uses a coarser sky grid than previously
described (5$^\circ$) and
only one spectral model table, the Band function medium spectrum defined above,
to which is added a pre-computed atmospheric scattering component that assumes
 the Earth is at the spacecraft nadir rather than calculating the atmospheric scattering
component at trigger time.  The first localization is produced by the FSW using the counts recorded in
the most significant data accumulation
on 17 time-scales ranging from 16 ms to 4.096 s, using data in the time
interval leading to the trigger and 
an additional 1.5~s of data accumulated following the trigger time, to find a
$\chi^2$ minimum in the on-board model rates table.  
The LAT
uses this location as a seed to allow a less stringent on-board trigger level than its usual all-sky on-board science algorithm.
The FSW produces further localizations if later data accumulations on time-scales from 16 ms to 4.096 s are more significant relative
to the background level 
than the initial accumulation at 1.5~s post-trigger, with these later localizations communicated both to the LAT and to the ground.  

In addition to the localizations produced
by the FSW, TDRSS is also used to transmit the most significant count rates above background on the 16~ms to 4.096 s accumulation intervals, called MAXRATES, 
with MAXRATES transmission occurring only if
the rates are more significant relative to the background level than previous MAXRATES calculations on any time-scale.
Finally, a background count-rate record is transmitted that contains the average count rates in each detector over a 16~s period prior to the trigger time and
separated from the trigger window by 3~s.
After reception on the ground,
MAXRATES and background packets, which also contain spacecraft position and attitude information, are ingested
 into the Burst Alert Processor (BAP) at NASA GSFC 
(or its backup at the GBM Instrument Operations Center in Hunstville).
The FSW-determined background rates are subtracted from the MAXRATES to give source rates that are compared to the models.    
A Ground-Auto localization is generated using the full-resolution model rates tables including 
an atmospheric response component that uses the true Earth position rather than assuming a zenith-pointed {\it Fermi} spacecraft. 
Both FSW and Ground-Auto localizations are communicated
as notices to the GRB Coordinates Network (GCN) if the statistical uncertainty is lower than previous FSW or
Ground-Auto notices.     

Over the next ten minutes, the FSW transmits via TDRSS a count rate time history
covering from 200~s prior to 450~s following the trigger time, and the BAP alerts the
GBM Burst Advocate (BA) to the presence of a trigger so that
a HitL location can be generated when the real-time data set is complete.
  Both the FSW and the Ground-Auto locations use single accumulations in intervals
from 16 ms to 4.096 s and can thus be considered peak-flux localizations. 
For the HitL localizations, the BA can select a source interval to produce a fluence localization that should 
in many cases yield a localization with a smaller statistical uncertainty. The
BA selects time intervals before and after the burst emission to fit with a polynomial 
of order up to 4 as a background model
that is subtracted from the observed counts in the source interval prior to the $\chi ^2$ minimization.

The three types of localizations are distributed without delay via GCN notices using email
 and socket connections.  FSW notices are sent out for all triggers and include a trigger classification (see \cite{meegan09} for 
a discussion of the FSW trigger classification procedure).  Ground-Auto notices are distributed
only if the FSW classified the trigger as a GRB at the time the MAXRATES packet was produced and the Ground-Auto localization 
passed an automated $\chi ^2$ quality test (i.e., the localization appears consistent with a distant point source based on the relative rates in the detectors).  
HitL positions are sent as Final Position notices only if the BA classifies the event as a GRB and the GRB has not been localized more
precisely by another instrument at the time of the HitL processing.
Table~\ref{tab:types} summarizes the types of locations, their reported statistical uncertainties, the typical latencies until the
first GCN notice is issued, and the number of notices of each type issued for a trigger classified by the FSW as a GRB.

\begin{table}
\centering
\begin{tabular}{|l|l|l|l|l|l|l|l|}
\hline
Notice Type& \multicolumn{2}{|c|}{Latency} & Type & \multicolumn{2}{|c|}{Error} & Number \\
 & Minimum & Typical & & Minimum & Typical &  \\
\hline
On-board (FSW) & 4 s$^1$ & 15-30 s & Peak Flux & 3 $^\circ$ & 8-15 $^\circ$ & 1-3  \\
Ground-Auto (GA) & 12 s & &  &  &  &   \\
GA 2008-2010 &  & 30-60 s& Peak Flux & 1 $^\circ$ & 5 $^\circ$ & 1-5  \\
GA 2011-2012 &  & 60-150 s& Peak Flux & 1 $^\circ$ & 5 $^\circ$ & 1-5  \\
GA current &  & 30-40 s& Peak Flux & 1 $^\circ$ & 5 $^\circ$ & 1-3  \\
Human-processed (HitL) & 19 min & 30-60 min & Fluence & 1 $^\circ$ & 3 $^\circ$ & 0-1 \\
\hline
\end{tabular}
\caption{Types of localization produced by GBM for GRB triggers. 
Localization uncertainties are 68\% CL statistical uncertainties. The
Ground-Automated notice latency has varied throughout the mission as described in the text, and each
configuration is listed separately. The Human-processed (HitL) notices are a new feature 
implemented in late 2011. {\footnotesize $^1$ This latency is lower than the time required
to activate the TDRSS link, suggesting the link was already active at trigger time}\label{tab:types}}
\end{table}

Latencies for the FSW notices have been stable since launch.  An update to the ground localization software in 2011 resulted in longer latencies
for the Ground-Auto notices owing to limitations of the BAP hardware and the processing in parallel of multiple MAXRATES packages.  An upgrade
of the BAP hardware occurred in 
2012 and the BAP processing software was also modified to reduce latencies and send notices only if the reported localization uncertainty is smaller
than previous notices for that trigger.  These changes have resulted in fewer, but more useful, 
Ground-Auto notices that are distributed more quickly. 
In 2011, the BAs began distributing the HitL position as a GBM notice, with latencies depending on data availability and BA response time.

\section{Comparison of GBM localizations with known GRB locations}
Between July 2008 and May 2013, GBM triggered on 203 GRBs that were well-localized by other instruments 
or by the IPN, with location uncertainties (68 \% confidence level)
smaller than $1^\circ$.  These 203 reference locations are listed in Appendix B, Table~\ref{tab:refs} .  The ground
localization software has been changed several times during the mission.
The current version of the code is 4.14g, for both the HitL and the Ground-Auto localizations, and this 
version is used 
for comparing the GBM locations with the 203 reference locations, so that the positions are recalculated using the
current version rather than using the GRB positions distributed via GCN notices and circulars.
It should also be noted that the HitL localizations used in this analysis were redone to ensure the background model and
source selection were uninfluenced by the known reference positions. 
On-board localization software has not changed since 2008 October 1 and 
the FSW locations are assessed using the 192 reference locations from GRBs that occurred after this date.

\begin{figure}
 \includegraphics{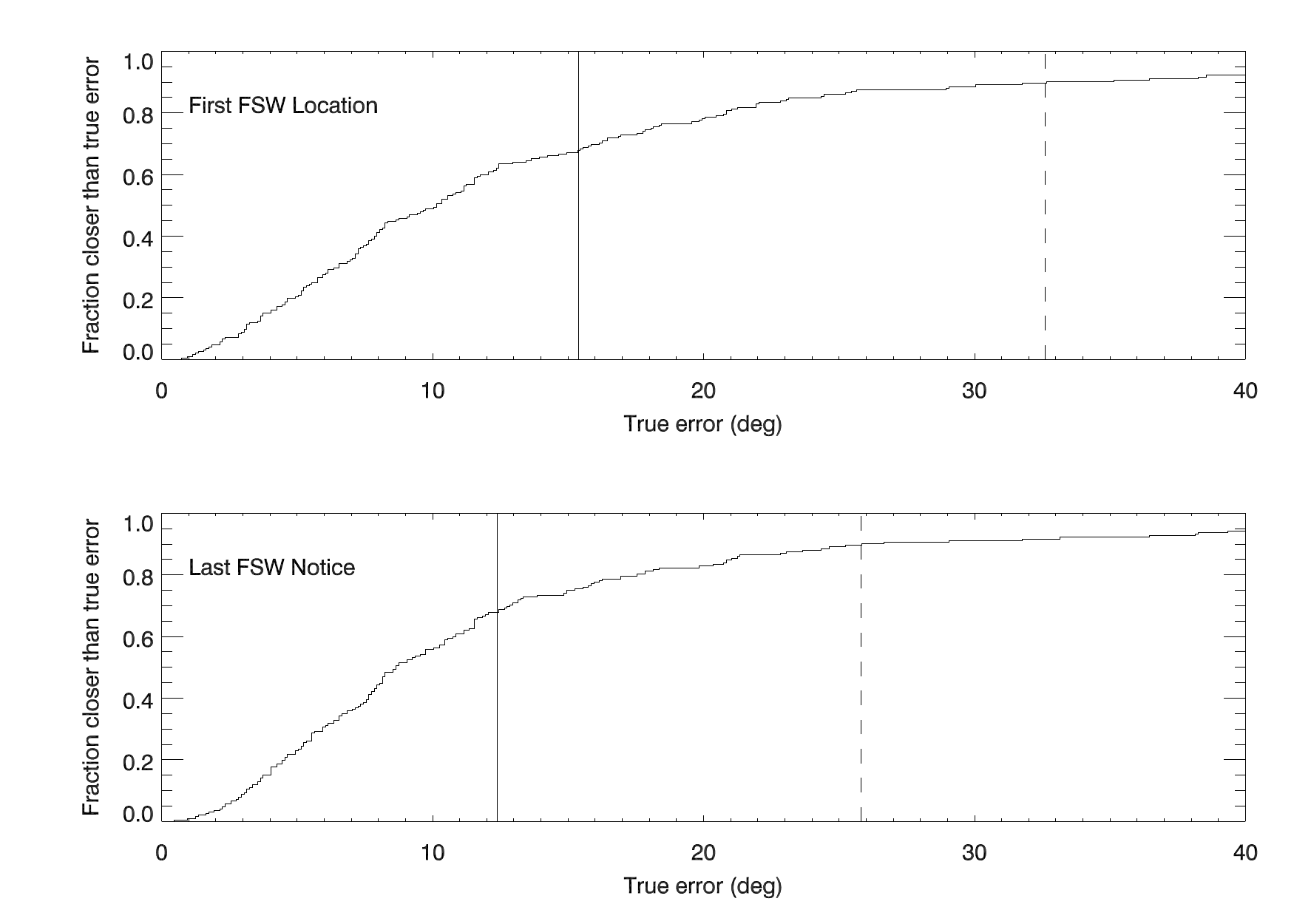}
 \caption{The histograms show the fraction of GBM FSW localizations
lying within a given offset (degrees) from the true positions for
 the initial (top) and final (bottom) FSW locations. The solid vertical lines
indicate the 68\% containment radius of 14.9$^\circ$ and 11.6$^\circ$, 
the dashed vertical lines the 90\% containment radius of 31.9$^\circ$ and 25.1$^\circ$.\label{fig:fsw_loc}}
 \end{figure}

The main purpose of the FSW locations is to alert the LAT to the occurrence of a GRB. This allows the LAT FSW
to adjust its on-board algorithm parameters using the temporal and spatial information from GBM. 
If the GRB is bright or spectrally hard enough, as determined by the GBM FSW, the FSW location is
also used to place the GRB near the center of the LAT Field-of-View (FoV) following an Automatic
Repoint Recommendation (ARR) issued by the GBM FSW to the spacecraft.  Because the FoV of the LAT
is $\sim 65^\circ$, the requirements on the FSW localization are loose, 20$^\circ$ uncertainty 
(68\% CL) with a goal of 15$^\circ$.
Figure~\ref{fig:fsw_loc}  shows the fraction of
FSW localizations within a given offset of the true location for the 192 reference locations.
The top panel shows the quality of the initial location, calculated at 1.5 s post-trigger. 
This can be compared in the bottom panel with the final FSW locations sent out as a GCN
notice, a set that includes the initial FSW localizations of GRBs for which only one FSW location was issued.
The vertical lines show that  68\% of the true positions are contained in a
14.9$^\circ$ [11.6$^\circ$] region for the initial [final] FSW locations, 
with 90\% contained within 31.9$^\circ$ [25.1$^\circ$].
This is sufficiently accurate for the LAT and the ARR process, and perhaps useful for the follow-up observer on the ground as an alert 
to begin slewing the telescope a few seconds before the more accurate Ground-Auto locations become available. 

\begin{figure}
 \includegraphics{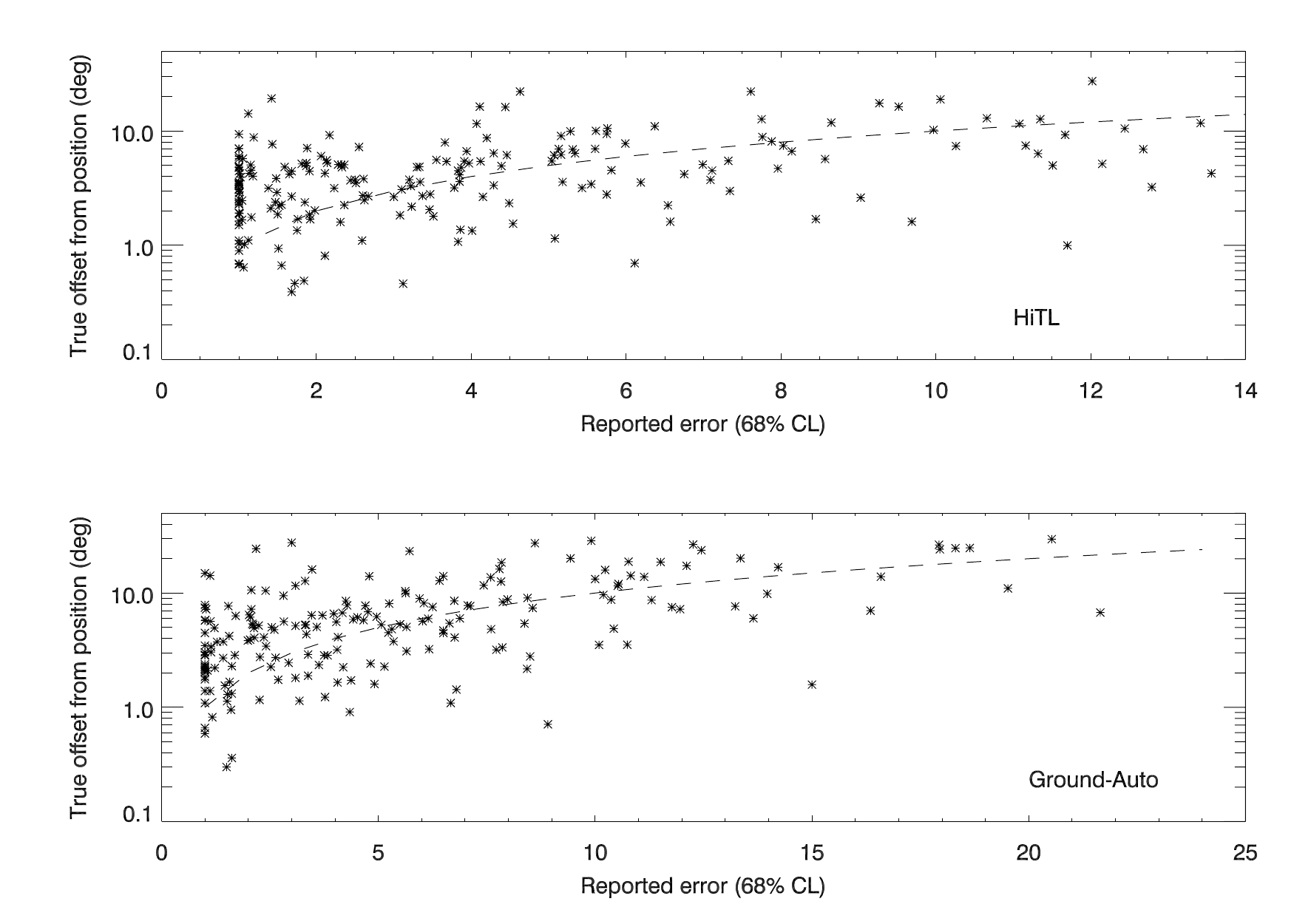}
 \caption{True offsets (degrees) from known positions for
  HitL (top) and Ground-automated (bottom) localizations as a function of the 68\% CL
  statistical uncertainties in the localization. The dashed line shows equality between the quantities.
A handful of Ground-Auto positions with uncertainties larger than 30$^\circ$ have been suppressed.\label{fig:locs_scatter}}
 \end{figure}

\begin{figure}
 \includegraphics{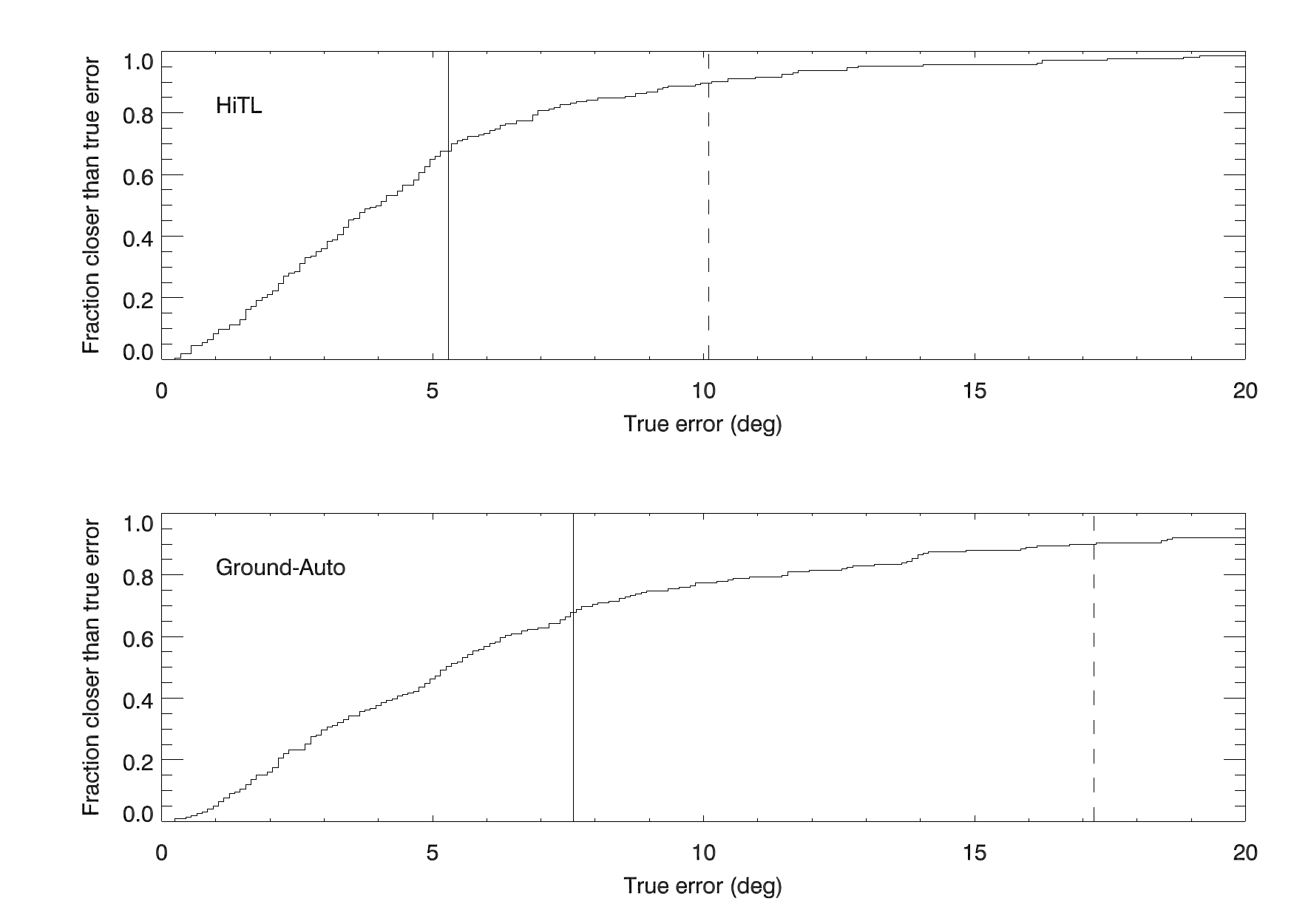}
 \caption{The histograms show the fraction of GBM localizations
lying within a given offset (degrees) from the true positions for
  HitL (top) and Ground-automated (bottom) positions. The solid vertical lines
indicate a 68\% containment radius of 5.3$^\circ$ [7.6$^\circ$], 
the dashed vertical lines the 90\% radius of 10.1$^\circ$ [17.2$^\circ$].\label{fig:locs_cum}}
 \end{figure}

\begin{figure}
 \includegraphics{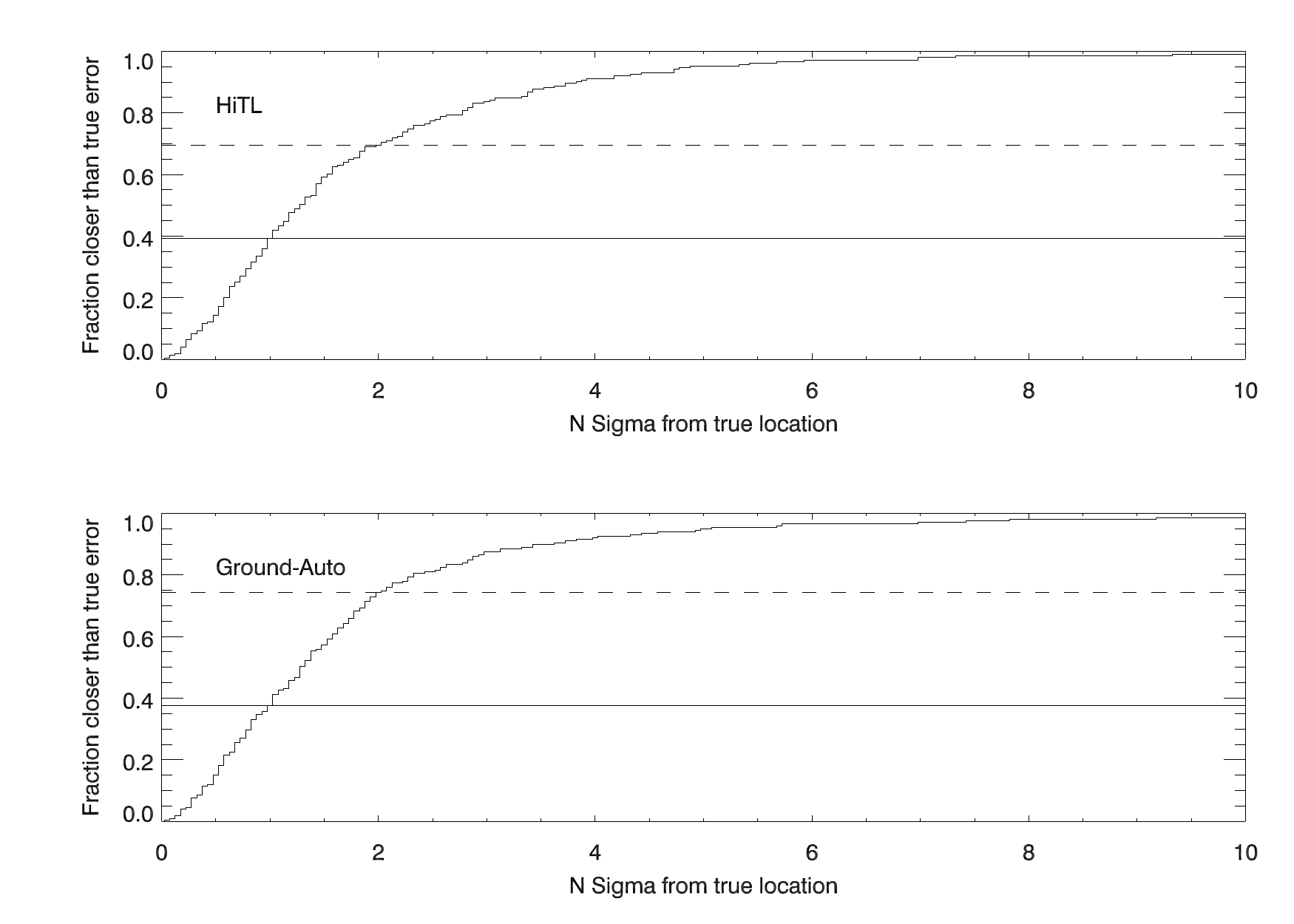}
 \caption{The histograms show the fraction of GBM localizations
lying within a given offset from the true positions for
  HitL (top) and Ground-automated (bottom) positions expressed as a multiple
of the 68\% 
statistical uncertainties. 
The solid horizontal lines
indicate the reported 1$\sigma$ containment radius
actually encompasses about 40\% of the events, with
the dashed lines showing about $70\%$ of events are contained
within the
 2$\sigma$ region.\label{fig:locs_cum_sigma}}
 \end{figure}

Figure~\ref{fig:locs_scatter} shows the true offset from the known
source position as a function of the reported statistical uncertainty for
both HitL (top) and Ground-Auto (bottom) localizations. Where
more than one possible Ground-Auto 
position exists, we use the last one produced by the BAP that would have been sent out as a notice.
Figure~\ref{fig:locs_cum}  shows the fraction of
GBM localizations within a given offset of the true location for HitL (top) and Ground-Auto locations.  The
vertical lines show that 68\% of the true positions are contained in a 5.3$^\circ$ [7.6$^\circ$] region around
the HitL [Ground-Auto] locations, with 90\% contained within 10.1$^\circ$ [17.2$^\circ$].
Although
the Ground-Auto positions appear significantly poorer, it can be seen from Figure~\ref{fig:locs_cum_sigma} that
when the offset to the true position is expressed as a multiple of the statistical uncertainty (68\% CL), the quality is similar.
This reflects the fact that the Ground-Auto locations are peak flux calculations that have fewer source
counts than the fluence HitL localizations, resulting in larger statistical uncertainties.  
The horizontal solid and dashed lines
show the fraction of localizations within 1 and 2$\sigma$. If the statistical uncertainties
reflected the total error, the 1$\sigma$ circles should contain 68\% of the true
source positions but they actually
contain 39\% [38\%] of the true source positions for the HitL and Ground-Auto locations, respectively, 
increasing to 70\% [74\%] for the $2\sigma$ regions.  This 
indicates that there is, in addition to the statistical uncertainty, a systematic component to the
localization error.  

 \begin{figure}
 \includegraphics{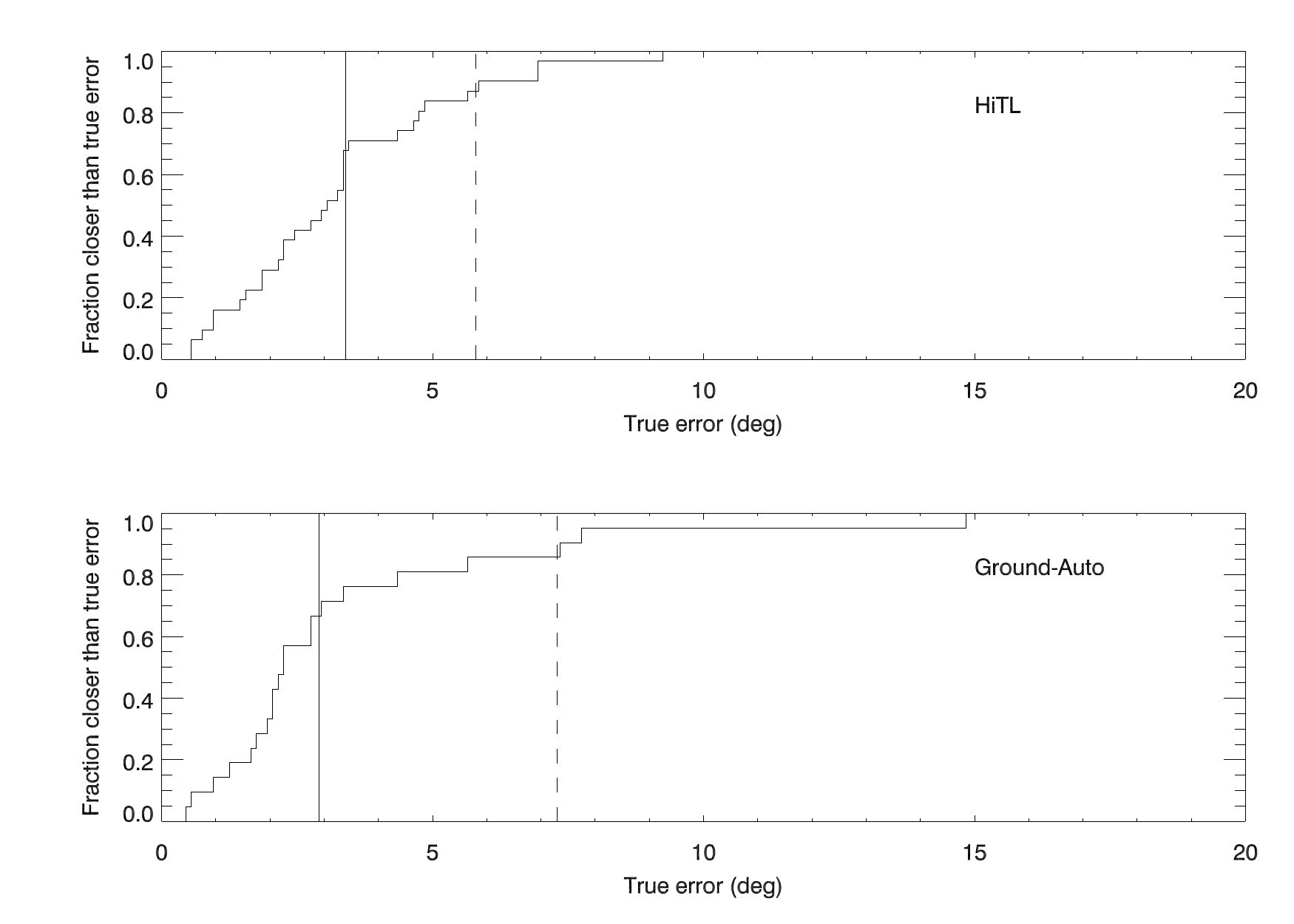}
 \caption{The histograms show the fraction of GBM localizations
lying within a given offset (degrees) from the true positions for
  HitL (top) and Ground-automated (bottom) positions with 
statistical uncertainties of 1$^\circ$. 
The solid vertical lines
indicate the 68\% containment radius (3.4$^\circ$ or 2.9$^\circ$), 
the dashed vertical lines the 90\% radius (5.8$^\circ$ or
7.3$^\circ$).
There are 31 GRBs in the HitL sample, 21 in the Ground-Auto sample.
The 68\% containment region for the 68\% containment radius for the HiTL
sample lies between 3.3 and $4.7^\circ$ with the 90\% containment
radius having a lower limit of $4.7^\circ$ and being unconstrained at the
upper end.
The 68\% containment region for the 68\% containment radius for the Ground-Auto 
sample lies between 2.2 and $5.6^\circ$ with the 90\% containment
radius having a lower limit of $4.3^\circ$ and being unconstrained at the
upper end.
\label{fig:locs_cum_1deg}}
 \end{figure}


Figure~\ref{fig:locs_cum_1deg} shows the fraction of precisely localized GRBs (statistical uncertainty equal to 1.0$^\circ$)
lying within a given distance of a known position for both HitL and Ground-Auto localizations.  


These distributions can serve as a guide to the follow-up observer wishing to concentrate on the brightest, best-localized
events without worrying too much about covering the whole uncertainty region, 
but wanting to know how often they will succeed as a function of
how much sky they are willing to tile. A more effective search strategy requires knowledge of the systematic uncertainty 
in order to probe the larger sky regions that contain most of the probability 
of the GRB arrival direction.  This is of particular interest for archival searches by the multi-messenger community with instruments that
do not require pointing a telescope to the source.  The sensitivity of such searches,
and of any upper limits obtained in the case of a null result,
depends on defining a source region narrow enough to reduce the background but that is still wide enough to capture the source.
In view of the connection between GRBs and core-collapse supernovae and the long time interval that can separate
the events, it is also essential to define GRB directions 
in order to connect or reject GRB associations with observed SNe events, as was done, for example,
 by \cite{soderberg2010} in the case of 
SN 2009bb.  This will become increasingly important in the era of the Large Synoptic
Survey Telescope (LSST) and the expected discovery of many more optical transients
in search of counterparts at other wavelengths.  

In the following section, we characterize the systematic uncertainty
associated with GBM localizations in order to be able to calculate the probability that any
region contains the actual source location.
We concentrate on the ground localizations 
that are most useful to devising a follow-up strategy although, for completeness, we will
briefly characterize the systematic uncertainty associated with the FSW localizations.
The reference sample includes the point source locations discussed above and annuli
obtained through triangulation by the IPN.
For analysis convenience we use only reference locations that may be considered
to be points and annuli that may be considered to be lines, with respect to
the GBM localization.  We include IPN annuli 
with $3\sigma$ half-widths
narrower than 0.8$^\circ$.
For some GRBs the IPN has multiple arcs that intersect to provide an accurate
location, with an intersecting region that has a 
corner-to-corner dimension of less than
$1.6^\circ$; in these cases we use the intersection closest to the GBM location
as a point location.
When the IPN has multiple arcs that do not provide an accurate location we
use only the narrowest arc in order not to overweight that GRB in the sample.
In addition to the 203 point locations described above,
134 GRBs from the IPN - GBM catalog \citep{ipn_catalog} between July 2008 and
July 2010 provided 244 annuli, shown in Appendix B, Table ~\ref{tab:ipns}.  After removing annuli that were wider than $1.6^\circ$ and collapsing intersecting annuli to point sources, 
this IPN catalog supplied 100 annuli and 9 additional point sources (from 18 intersecting annuli). The $1.6^\circ$ acceptance limit for the
IPN annuli ensures that only annuli with dimensions smaller than the smallest GBM localization uncertainty are used in the reference sample.
Wider annuli would not influence the model fits because their contribution to the constraints on the model parameters would be weighted by 
the large uncertainty in the dimension (width) that is
used to evaluate the best-fit parameters and the goodness-of-fit of the model, but the reference sample would appear larger than the true useful
reference sample.
These criteria then provide a reference sample with $N_{\rm point} = 212$ point locations
and $N_{\rm arc} = 100$ arcs.

\begin{figure}
\vbox{
\center{
\hbox{
 \includegraphics[width=9cm]{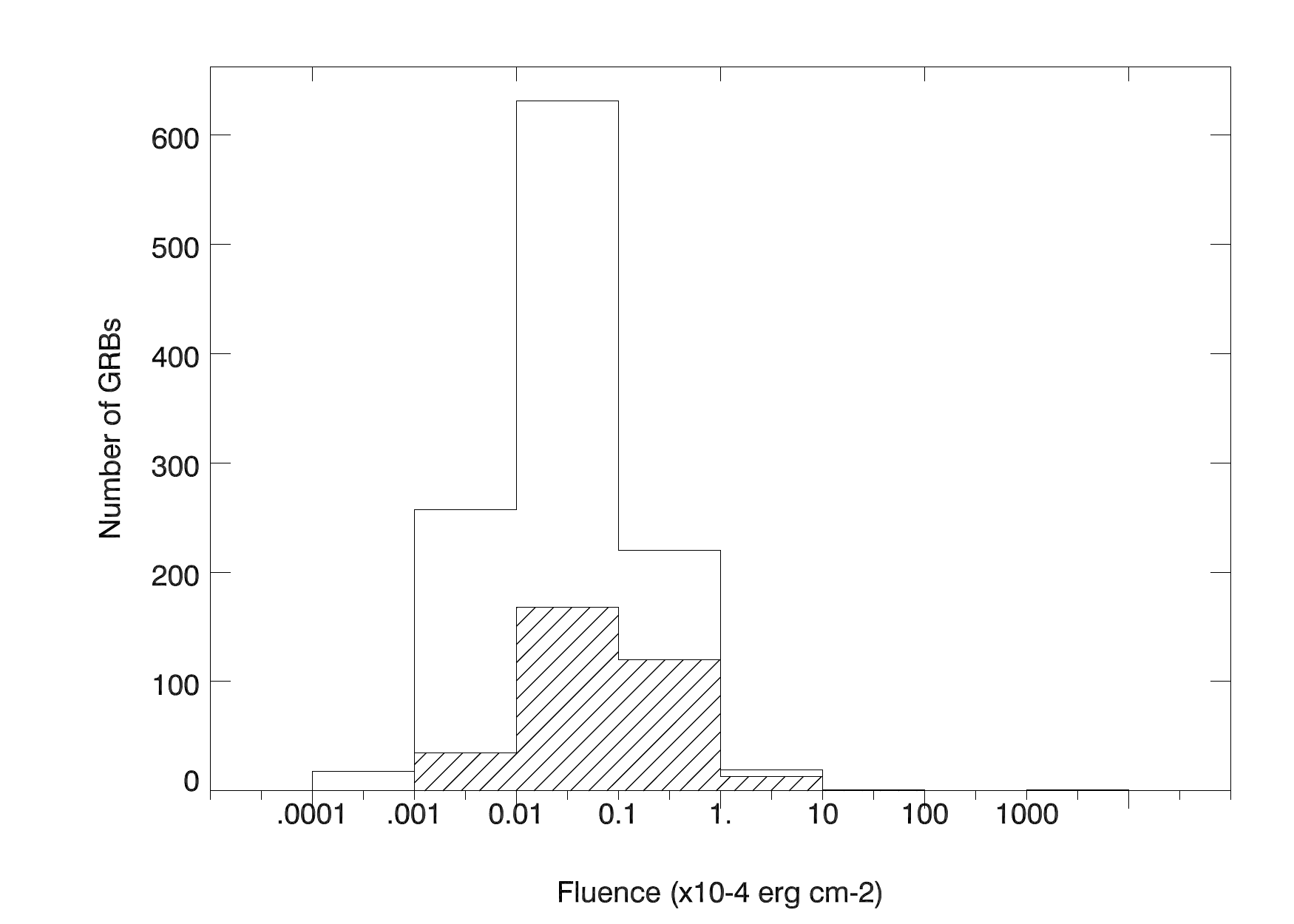}
 \includegraphics[width=9cm]{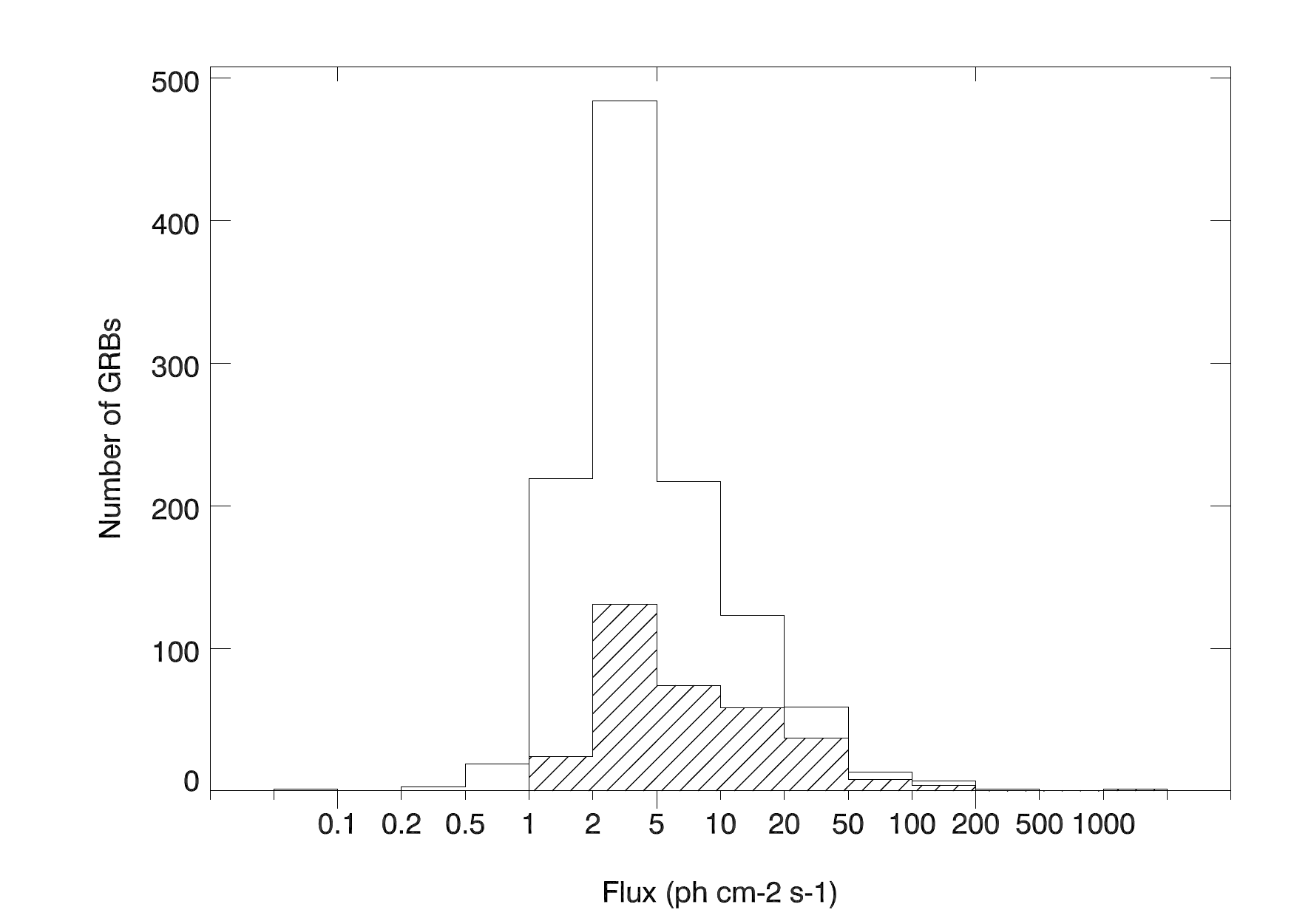}}}
\center{
 \includegraphics[width=9cm]{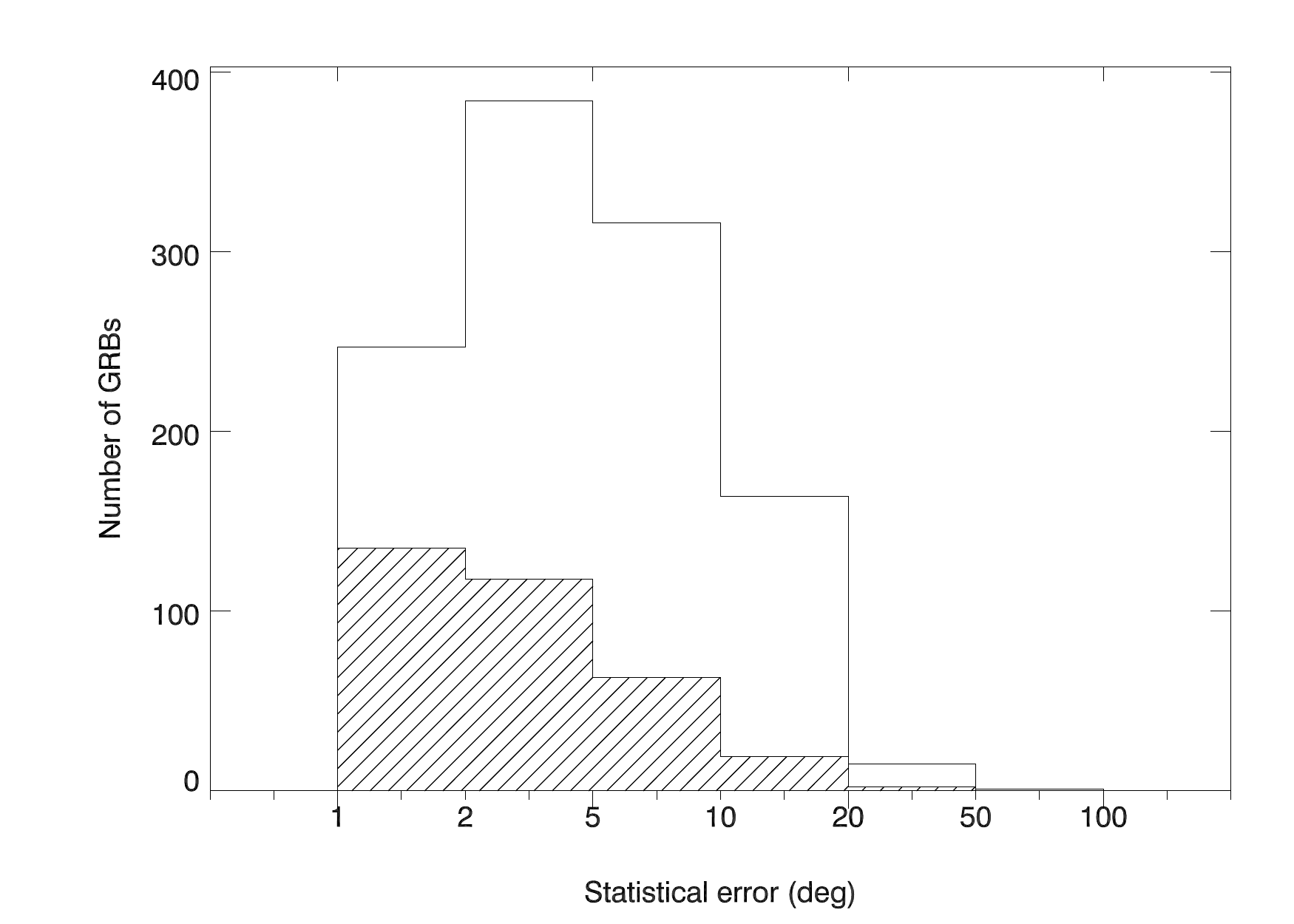}}}
 \caption{Comparison of the properties of GRBs in the reference sample (hashed) to 
those of the general population of GRBs detected by GBM (solid). 
The panels show (a) fluence (b) 1-s peak flux 
between 10 - 1000 keV (c) 1-$\sigma$ statistical uncertainty on the 
HitL localization.
\label{fig:sample_comp}}  
 \end{figure}

In Figure ~\ref{fig:sample_comp}, we compare
the properties of our reference sample to the population of GRBs detected over the same period of time.
Although the GRBs in our reference sample are, on average, brighter than other GRBs detected by GBM,
with more source counts leading to smaller statistical localization errors, they are more representative of
the overall population of GRBs detected by GBM than the reference sample in ~\citet{briggs99} is of the
overall GRB population detected by BATSE.  The reference GRB population in this work is drawn from 
experiments with higher fluence thresholds (IPN, {\it Fermi}-LAT) but also from experiments
with equal or lower fluence sensitivity ({\it Swift}-BAT, INTEGRAL), whereas the reference sample in
\citet{briggs99} comprised only GRBs bright enough for detection by the IPN.

\section{Investigating the systematic uncertainty associated with GBM localizations}
We use a Bayesian approach to find a model characterizing the systematic uncertainties in GBM localizations.  
This approach was used by \cite{briggs99} to show that the systematic uncertainties in BATSE GRB localizations were better fit using
a model that contained most of the probability in a core distribution that peaked at 1.6$^\circ$, 
with a larger uncertainty in an extended tail, than by a single component. 

The GBM localization program estimates a statistical location uncertainty,
$\sigma_{\rm stat}$.  The localization errors are larger due
to systematic errors, so we
represent the total location uncertainty as $\sigma_{\rm tot}^2 =
\sigma_{\rm stat}^2 + \sigma_{\rm sys}^2$.
The errors are assumed to be azimuthally symmetric.
The models are based on the Fisher probability density function,
which has been called the Gaussian distribution on the sphere (Fisher et al. 1987):
\begin{equation}
p_{\rm F}(\gamma) \; d \Omega =
\frac{\kappa}{2 \pi ({\rm e}^{\kappa} - {\rm e}^{-\kappa})}
      {\rm e}^{\kappa \cos{\gamma}} \; d \Omega,
\end{equation}
where $\gamma$ is the angle between the measured and true location,
$\kappa$ is termed the concentration parameter and $d \Omega$ is solid angle.
Considering $\sigma_{\rm tot}$ to be the radius  of the circle containing
68\% of the total probability, integrating eq.~1 relates $\kappa$ and $\sigma_{\rm tot}$ in {\it radians} \citep{briggs99}:
\begin{equation}
\kappa = \frac {1} {(0.66 \sigma_{\rm tot})^2}.
\end{equation}
We find equation 2 works well over a broad range of
$\sigma_{\rm tot}$ and use it for all values in our sample.

For comparison with point reference locations, we denote the separation between a GBM localization
and the reference location as $\gamma$, while for an arc we denote the
separation beween the GBM localization and the closest point on the center-line of the annulus
as $\rho$.
Of course, the true separation might be larger than $\rho$.
Our aim is to develop and test models for the GBM localization probability
density function $p_\gamma(\gamma)$.
We assume the GBM localization probability $p_\gamma(\gamma)$ is a Fisher function or the sum
of two Fisher functions:
\begin{equation}
p = f p_{\rm F}(\gamma_1)  + (1-f) p_{\rm F}(\gamma_2).
\end{equation}
For these cases, the probability density function $p_\rho(\rho)$ is  known analytically \citep{briggs99}.

We use Bayesian
model comparison to test models for the systematic error \citep{loredo,sivia}.
Bayesian model comparison is based on the likelihood, which is the product
of the probabilities of the observed offsets:
        \begin{equation}
            \mathcal{L} = \prod_{i} p_\gamma (\gamma_i)  \prod_{j} p_\rho (\rho_j).
        \end{equation}
The equation shows the dependence of the likelihood $\mathcal{L}$ on the offsets
$\gamma_i$, $i=1,..., N_{\rm point}$ and $\rho_j$, $j=1,..., N_{\rm arc}$; $\mathcal{L}$ also
depends on the error model and is a function
of the error model parameters through the functions $ p_\gamma$ and $p_\rho$.
The reference data are of two types, point locations and arcs, disparately
testing the quality of the GBM localizations.  The Bayesian model comparison
naturally handles the difference in evidence since both types of reference data are included in the
likelihood via probabilities.

A  complicated model with many parameters  may have improved likelihood because
the model is better or because the additional parameters allow the model to conform to statistical
fluctuations.  The question is, is the improvement in likelihood
sufficient that we should believe in the more complicated model?
Bayesian model comparison handles this problem by penalizing
models with additional parameters with Occam's factors $F$.
Assuming that a reasonable range, or prior, for the parameter $\lambda_k$
is from $\lambda_k^{\rm min}$ to $\lambda_k^{\rm max}$, that the
model fitting has estimated the parameter uncertainty as
$\sigma_{\lambda_k}$ and that the likelihood function is approximately
Gaussian, the Occam's factor for parameter $\lambda_k$ is \citep{sivia}: 
\begin{equation}
F_k = \frac{ \sigma_{\lambda_k} \sqrt{2\pi} }
{\lambda_k^{\rm max} - \lambda_k^{\rm min}}.
\end{equation}
We consider all of the models equally plausible (i.e., identical priors),
so that the Odds ratio $ O_{\rm B/A}$ by which one should favor model
B over model A is
        \begin{equation}
              O_{\rm B/A} =  \frac {P({\rm B})} {P({\rm A})}
      =     \frac {  \mathcal{L}({\rm B}) \times \prod F_b^B }
                { \mathcal{L}({\rm A}) \times \prod F_a^A  }.
        \end{equation}

Instead of listing the Odds ratios $ O_{\rm B/A}$ for every combination of
models $A$ and $B$ we calculate in Tables \ref{tab:single} to \ref{tab:fsw_rock} the quantities
\begin{equation}
\log_{10} [ \mathcal{L}({\rm M}) \times \prod F_k^M ].
\end{equation}
The base-10 logarithm of the Odds factor of two models may be obtained
by differencing these values.

The following models were tested (where we use the term Gaussian to denote the Fisher function):
\begin{itemize}
\item
A single component fit by a Gaussian. This model has a single parameter, the
peak position of the Gaussian.
\item
A single component modeled by a Gaussian,  split into 2 populations depending
on the hemisphere containing the GRB position, in spacecraft coordinates. One model each for the X, Y and Z hemispheres splits
the GRBs into positive and negative hemispheres.
These models have two parameters: the peak of the Gaussian for bursts in each hemisphere.
\item
To explore the effects of the symmetry of the {\it Fermi} spacecraft on
the GBM localization accuracy, the samples were also divided according to quadrant, with GRB positions within the $\pm$X
quadrants separated from those localized within the $\pm$Y quadrants.  
These models have two parameters: the peak of the Gaussian for bursts in each quadrant set.
\item
A core-plus-tail modeled by two Gaussians.  This model has three parameters: the peak position of each Gaussian and the fraction in
the core.
\item
A core-plus-tail modeled by two Gaussians, split into 2 populations depending
on the hemisphere of the GRB position, in spacecraft coordinates (one model each for the X, Y, and Z hemispheres).  
Each model has six parameters: the peaks of the two Gaussians and the fraction in
the core, for each hemisphere.
\item
A core-plus-tail modeled by two Gaussians, split into two sets of quadrants, one containing the GRBs
 located within the $\pm$X quadrants, the other within the $\pm$Y quadrants. 
This model, like the hemisphere models, has six parameters.
\end{itemize}

\begin{table}
{\footnotesize
\begin{tabular}{|l|l|l|c|c|l|l|}
\hline
\multicolumn{7}{|c|}{Single-Component Gaussian} \\
\hline
Type & \multicolumn{2}{|c|}{Number GRB} & Peak & Error & Log$_{10}$ & Log$_{10}$ \\
& Point & Annuli & $^\circ$ & $^\circ$ & Likelihood  & Odds Factor \\
\hline
Ground-Auto & 208 & 100 & 6.62 & 0.29 & 155.2 & 154.3 \\
HitL 4.14g & 212 & 100 & 5.15 & 0.22 & 224.6 & 223.6 \\
HitL 4.13 & 211 & 100 & 6.23 & 0.26 & 192.6 & 191.7 \\
\hline
\end{tabular}
\caption{Single-component fits to systematic uncertainties on GBM localizations of different types.}\label{tab:single}}
\end{table}

\begin{table}
{\footnotesize
\begin{tabular}{|l|l|l|c|c|c|c|c|c|l|l|}
\hline
\multicolumn{11}{|c|}{Core + Tail (2 Gaussians)}  \\
\hline
Type & \multicolumn{2}{c}{Number GRB} & Core & Error & Core & Error & Tail & Error & Log$_{10}$ & Log$_{10}$ \\
& Point & Annuli &  $^\circ$ &  $^\circ$ &  \% & \%  & $^\circ$ & $^\circ$ & Likelihood  & Odds Factor \\
\hline
Ground-Auto & 208 & 100 & 3.72 & 0.34 & 80.4 & 5.3 & 13.7 & 1.7 &  174.1 & 171.8 \\
HitL 4.14g & 212 & 100 & 3.71 & 0.24 & 90.0 & 3.5 & 14.3 & 2.5 &  241.3 & 238.9 \\
HitL 4.13 & 211 & 100 & 3.57  & 0.32 & 79.8 & 5.3 & 12.7 & 1.5 & 213.5 & 211.1  \\
\hline
\end{tabular}
\caption{Core-plus-tail fits to systematic uncertainties on GBM localizations of different types.}\label{tab:corept}}
\end{table}

Tables~\ref{tab:single} and \ref{tab:corept}
show the results for the most basic single-component and core-plus-tail
fits to the systematic uncertainties on the Ground-Auto, and HitL localizations. In addition to the current version
of the localization software, 4.14g,
fits for version 4.13 of the HitL localization are 
included in order to characterize the localizations in the first pair of GRB catalogs and part of the second
pair of GRB catalogs \citep{paciesas_1grb, goldstein_1grb, azk_2grb, dg_2grb}, which
used this older version. The main change implemented in 4.14g was the selection of a model spectrum based on the
$\chi^2$ values of the
best-fit location using all three model spectra rather than {\it a priori} using three bands of hardness ratio values
to classify the GRB as soft, medium, or hard and finding the $\chi^2$ minimum only in the single corresponding model rates table.
Looking at the difference between the odds factors between
the single-component fits in Table~\ref{tab:single} and the core-plus-tail fits in Table~\ref{tab:corept}, 
it can be seen that a core-plus-tail model is favored over a single component model by factors of
$10^{17}$, $10^{16}$, and $10^{19}$ for the Ground-Auto, HitL 4.14g, and HitL 4.13, respectively.  
The fraction in the core 
for the HitL localizations is higher
using the more recent code, consistent with Version 4.14g being more robust, although the core values agree
within errors, and are in fact slightly lower using the older code.  

Results for single-component models exploring the effects of GRB location  
in spacecraft coordinates on the systematic error are
shown in Appendix C. Table~\ref{tab:single_geom} shows that 
parameter values are similar for the systematic errors of GRBs in each of the hemispheres.  
There is no consistent 
statistical preference 
among the three localization types 
for models based on GRB position in spacecraft coordinates although individual localization types
show slight preferences for hemisphere- or quadrant- dependent models. 
Owing to the placement of the GBM NaI detectors, which maximizes
sensitivity in the +Z hemisphere to optimize coverage of GRBs in the LAT FoV, GRBs in the +Z
hemisphere are more plentiful, viewed by more detectors, and might be expected to have more accurate localizations.
The weak sensitivity of the systematic error
to GRB position along the Z-axis suggests that any such effect
is encompassed with a larger statistical error for the bursts viewed with fewer detectors in the -Z hemisphere.  
If we look at the cumulative fraction of GRBs lying within a given offset of the true location, 
we find that for HitL 4.14g the 68\% [90\%] containment radius is 5.1$^\circ$ [9.9$^\circ$] for GRBs
localized in the +Z hemisphere versus 5.3$^\circ$ [10.9$^\circ$] for those in the -Z hemisphere.
This difference is larger for version 4.13: 6.1$^\circ$ [11.1$^\circ$] in the +Z hemisphere
versus 6.8$^\circ$ [15.0$^\circ$] in the -Z hemisphere, consistent with the results shown in Table~\ref{tab:single_geom}.  
These numbers also suggest the overall
quality of the 4.14g locations is higher than the older version and fewer outliers are produced.
The odds ratios for all of the single-component models are much
lower than the core-plus-tail model in Table~\ref{tab:corept}, which is thus greatly preferred over
any of the single-component models. 

Results for core-plus-tail models that explore the effect of GRB position in different 
spacecraft hemispheres are described in Appendix D. 
Table~\ref{tab:corept_hemi} shows that 
localizations using HitL 4.14g are fit 
with lower parameter values for GRBs in the +Z hemisphere, although this model is
not preferred statistically over the simpler, all-sky core-plus-tail model in Table~\ref{tab:corept}. Other 
hemisphere-dependent core-plus-tail models did not yield significantly different parameters in the
two hemispheres, nor were they statistically favored.

\begin{table}[!h]
{\footnotesize
\begin{tabular}{|l|l|l|l|c|c|c|c|c|c|l|l|}
\hline
\multicolumn{12}{|c|}{Core + Tail (2 Gaussians) with quadrant dependence}  \\
\hline
Type & Quadrants & \multicolumn{2}{c}{Number GRB} & Core & Error & Core & Error & Tail & Error & Log$_{10}$ & Log$_{10}$ \\
& & Point & Annuli &  $^\circ$ &  $^\circ$ &  \% & \%  & $^\circ$ & $^\circ$ & Likelihood  & Odds Factor \\
\hline
Ground-Auto & All-sky & 208 & 100 & 3.72 & 0.34 & 80.4 & 5.3 & 13.7 & 1.7 &  174.1 & 171.8 \\
Ground-Auto & $\pm$ X & 208 & 100 & 4.10 & 0.48 & 75.4 & 8.2 & 13.1 & 2.1 &  & \\
& $\pm$ Y &  &  & 3.06 & 0.61 & 80.6 & 8.9 & 12.9 & 2.9 &  175.0 & 171.7 \\
\hline
HitL 4.14g & All-sky &  212 & 100 & 3.71 & 0.24 & 90.0 & 3.5 & 14.3 & 2.5 &  241.3 & 238.9 \\
HitL 4.14g & $\pm$ X & 212 & 100 & 4.30 & 0.38 & 90.3 & 5.9 & 15.0 & 4.8 &   & \\
 & $\pm$ Y &  & & 3.34 & 0.26 & 92.0 & 3.5 & 14.5 & 3.5 &  242.6 & 238.7 \\
\hline
HitL 4.13 & All-sky & 211 & 100 & 3.57  & 0.32 & 79.8 & 5.3 & 12.7 & 1.5 & 213.5 & 211.1  \\
HitL 4.13 & $\pm$ X & 211 & 100 & 4.22 & 0.87 & 72.8  & 15.5  & 11.5 & 2.4 & &  \\
 & $\pm$ Y &  &  & 3.24  & 0.35 & 89.4 & 5.1 & 16.8 & 3.3 & 216.2 & 213.0 \\
\hline
\end{tabular}
\caption{Quadrant-dependent core-plus-tail fits to systematic uncertainties on GBM localizations.}\label{tab:corept_quad}}
\end{table}

Table~\ref{tab:corept_quad} shows
the model fit parameters for localizations grouped according to GRB position
within the $\pm$X and $\pm$Y quadrants.  
Unlike the grouping into hemispheres, the differences in parameter values for quadrant-dependent
models for all three types of localization cannot
be explained by juggling events from the core to the tail.  Instead, the $\pm$Y quadrant GRBs are more likely
to be localized with a systematic error in the core, and the value of the systematic error in the core is smaller than for
events in the $\pm$X quadrants.  The odds factors for these models
can be compared with the all-sky core-plus-tail model repeated in line 1 of the table.  
 This 6-parameter model is favored only for HitL 4.13, which had
a larger population of outliers in the tail than HitL 4.14, but the parameter values are suggestive that this 6-parameter
model should be favored over the simpler 3-parameter core-plus-tail model with a larger sample or more judiciously
chosen quadrants. 

\begin{figure}
 \center{
\includegraphics[width=5in]{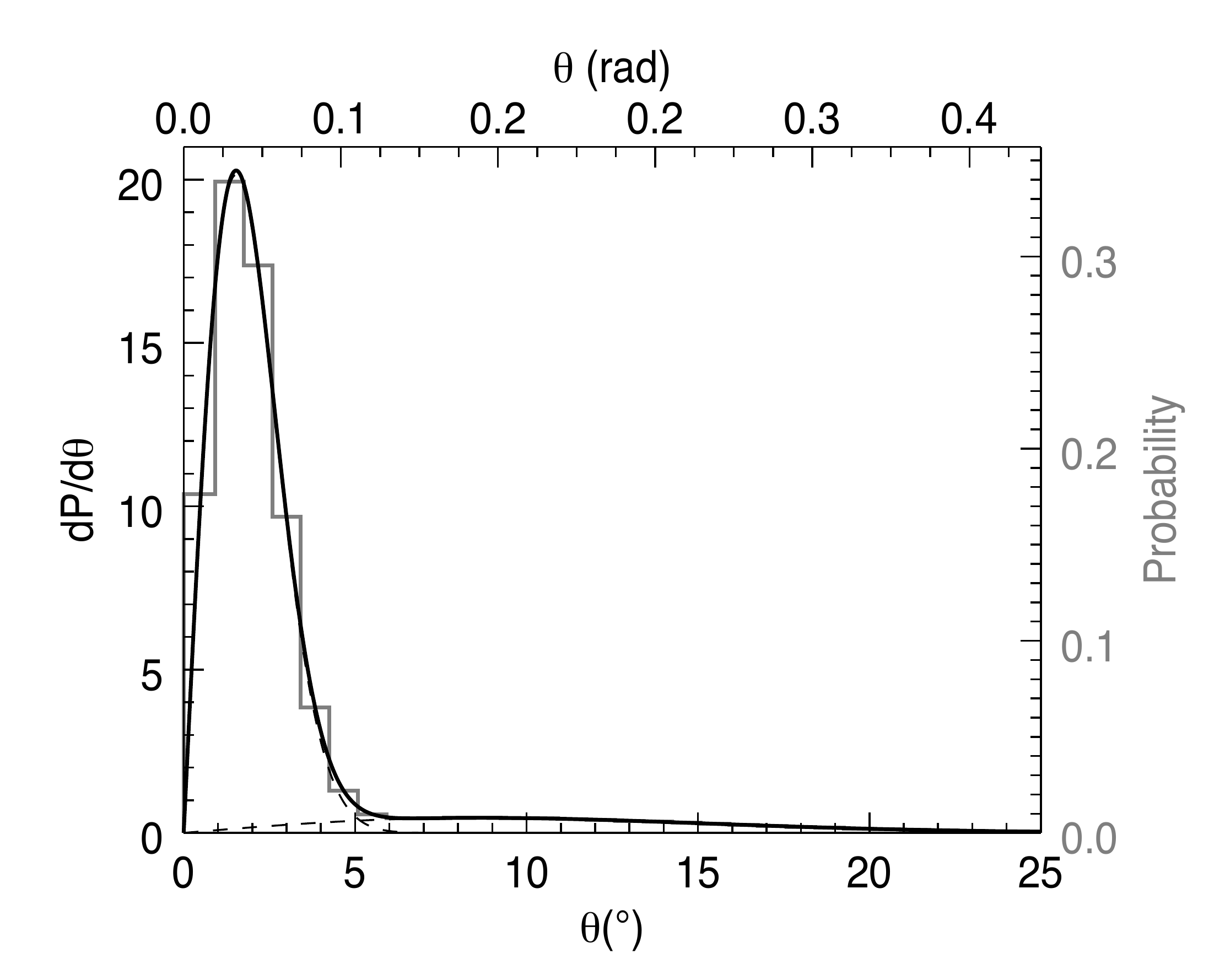}}
 \caption{The solid curve shows
the sum of the two components in the core and tail of a model representing
the total 68\% confidence level systematic uncertainty for GRBs localized
in the $\pm Y$ quadrants in the spacecraft coordinate frame.  The dashed curves 
show the individual components for the core and tail of the model.
The histogram shows the probability of a systematic
error for a given angular bin. 
\label{fig:sysgauss}}  
 \end{figure}

Table~\ref{tab:corept_quad_slide} displays the quadrant-dependent core-plus-tail model parameters and 
odds ratios
with sliding quadrant azimuth ranges. A larger sample of reference locations will, in the future, allow the quadrant ranges to
vary as free parameters but a model with these extra free parameters
is too complicated to converge with the current sample.  The core-plus-tail models are significantly preferred
over single Gaussians so that optimizing the quadrant azimuth ranges with a single Gaussian model is not explored. 
  Entry 1 is the standard all-sky core-plus-tail model, which was
found to represent the data better than any of the single-component models we explored.
Entry 2 has equal azimuth ranges in each quadrant.  Entry 3 explores the effect of making the $\pm$Y 
quadrant bigger.  This provides no improvement and moves the quadrant core parameter values closer to each other. Narrowing
the $\pm$Y quadrant produces fits that increasingly differentiate between the quadrants, with increasingly favorable odds
factors exceeding the simple 3-parameter model.
When the $\pm$Y quadrant becomes too small to contain enough events to constrain the parameters, the odds factor
once more decreases.  The model that maximizes the odds factor, entry 7, also differentiates the most between the quadrants,
with small parameter errors and close to $90\%$ of GRBs in the core in each quadrant set. It is preferred over the all-sky
core-plus-tail model by a factor of 16.
This set of parameters defines $\pm$Y quadrants with $45^\circ$
azimuth in each quadrant and lower systematic errors for GRBs localized in 
these quadrants than for those
in the $135^\circ \pm$X quadrants.
This quadrant definition
also produces larger odds factors for the Ground-Auto localizations than the core-plus-tail model 
that has equal $90^\circ$ quadrants, with
the Ground-Auto providing an independent data set from the same GRB sample. This quadrant-dependent core-plus-tail
model for the systematic uncertainty associated with GBM GRB localizations is illustrated in Figure \ref{fig:sysgauss}. The figure
shows how the probability of the systematic uncertainty changes
as a function of angular offset from a GBM GRB localization for GRBs localized in the $\pm Y$ quadrants.

\begin{table}[!h]
{\footnotesize
\begin{tabular}{|c|l|l|c|c|c|c|c|c|l|l|}
\hline
\multicolumn{11}{|c|}{Core + Tail (2 Gaussians) varying the azimuth boundaries of the quadrants}  \\
\multicolumn{11}{|c|}{Version 4.14g of the HitL code with 212 reference locations and 100 IPN annuli}  \\
\hline
 \# & Quadrants & Azimuth ranges & Core & Error & Core & Error & Tail & Error & Log$_{10}$ &  Log$_{10}$ \\
 & & & $^\circ$ &  $^\circ$ &  \% & \%  & $^\circ$ & $^\circ$ & Likeli-  & Odds \\
 & & &  &   &   &   &  &  & hood  & Factor \\
\hline
1 & All-sky & 0 - 360$^\circ$  & 3.71 & 0.24 & 90.0 & 3.5 & 14.3 & 2.5 &  241.3 & 238.9 \\
\hline
2 &  $\pm$ X & 315 - 45$^\circ$ and 135 - 225$^\circ$ &  4.30 & 0.38 & 90.3 & 5.9 & 15.0 & 4.8 &   & \\
&  $\pm$ Y & 45 - 135$^\circ$ and 225 - 315$^\circ$ &  3.34 & 0.26 & 92.0 & 3.5 & 14.5 & 3.5 &  242.6 & 238.7 \\
\hline
3 &  $\pm$ X & 320 - 40$^\circ$ and 140 - 220$^\circ$ &  4.23 & 0.52 & 87.2 & 7.7 & 14.1 & 4.8 &   & \\
& $\pm$ Y & 40 - 140$^\circ$ and 220 - 320$^\circ$ &  3.47 & 0.25 & 93.6 & 3.3 & 15.5 & 3.6 &  242.6 & 238.8 \\
\hline
4 & $\pm$ X & 305 - 55$^\circ$ and 125 - 235$^\circ$ &  4.25 & 0.32 & 92.6 & 4.6 & 15.9 & 4.7 &   & \\
&  $\pm$ Y & 55 - 125$^\circ$ and 235 - 305$^\circ$ &   3.16 & 0.45 & 90.2 & 4.8 & 14.8 & 3.6 &  242.6 & 238.8 \\
\hline
5 &  $\pm$ X & 300 - 60$^\circ$ and 120 - 240$^\circ$ &  4.32 & 0.29 & 93.4 & 4.0 & 16.2 & 4.9 &   & \\
& $\pm$ Y & 60 - 120$^\circ$ and 240 - 300$^\circ$ &  2.59 & 0.40 & 86.6 & 6.1 & 13.5 & 3.2 &  243.9 & 240.1 \\
\hline
6 & $\pm$ X & 295 - 65$^\circ$ and 115 - 245$^\circ$ &  4.17 & 0.28 & 91.3 & 3.9 & 15.5 & 3.6 &   & \\
& $\pm$ Y & 65 - 115$^\circ$ and 245 - 295$^\circ$ &  2.42 & 0.38 & 89.3 & 5.7 & 13.2 & 3.9 &  244.2 & 240.2 \\
\hline
7 & $\pm$ X & 292.5 - 67.5$^\circ$ and 112.5 - 247.5$^\circ$ &  4.17 & 0.28 & 91.8 & 3.9 & 15.3 & 3.6 &   & \\
& $\pm$ Y & 67.5 - 112.5$^\circ$ and 247.5 - 292.5$^\circ$ &  2.31 & 0.39 & 88.4 & 6.4 & 13.2 & 3.8 &  244.2 & 240.3 \\
\hline
8 &  $\pm$ X & 290 - 70$^\circ$ and 110 - 250$^\circ$ &  4.12 & 0.28 & 91.7 & 3.9 & 14.9 & 3.6 &   & \\
& $\pm$ Y & 70 - 110$^\circ$ and 250 - 290$^\circ$ &  2.35 & 0.43 & 87.0 & 7.0 & 12.7 & 4.0 &  243.7 & 239.9 \\
\hline
9 & $\pm$ X & 285 - 75$^\circ$ and 105 - 255$^\circ$ &  3.96 & 0.26 & 89.9 & 3.8 & 15.3 & 2.9 &   & \\
& $\pm$ Y & 75 - 105$^\circ$ and 255 - 285$^\circ$ &  2.64 & 0.54 & 90.7 & 10.6 & 9.1 & 4.8 &  243.0 & 239.4 \\
\hline
\end{tabular}
\caption{Effect of varying the quadrant boundaries when assessing the quadrant-dependent 
core-plus-tail fit to the systematic uncertainty on GBM localizations.}\label{tab:corept_quad_slide}}
\end{table}

\subsection{Flight Software systematic localization uncertainties}
Both the Initial and Final FSW locations had sufficiently
 distant outliers ($> 90^\circ$) that most single component fits did not
converge.  Removing the five worst localizations allowed the fitting of several single component models for the Initial
locations, shown in Table~\ref{tab:fsw_init}.  The IPN annuli were also removed as it could not be determined
for these reference locations which were the worst outliers and the fits failed to converge when they were included.  
The single-component
fits for the FSW Initial locations are presented because these were the only fits that converged, and the parameter values
may be useful, but it should be noted that these fits omit the five worst localizations.

\begin{table}
{\footnotesize
\begin{tabular}{|l|l|l|c|c|l|l|}
\hline
\multicolumn{7}{|c|}{FSW Initial Locations: Single Component models across hemispheres} \\
\hline
Model & \multicolumn{2}{|c|}{Number GRB} & Peak & Error & Log$_{10}$ & Log$_{10}$ \\
& Point & Annuli & $^\circ$ & $^\circ$ & Likelihood  & Odds Factor \\
\hline
Single & 187 & 0 & 15.4 & 0.8 & 3.7 & 3.2  \\
Hemisphere +X & 187 & 0 & 12.2 & 1.2 &   &   \\
-X &  &  &  17.2 & 1.2 & 5.5 & 4.9  \\
Hemisphere +Y & 187 & 0 & 12.7 & 1.0 &   &   \\
-Y &  &  &  18.2 & 1.4 & 6.0 & 5.4  \\
Quadrants $\pm$X & 187 & 0 & 19.7 & 1.6 &   &   \\
$\pm$Y &  &  &  12.5 & 0.9 & 7.5 & 6.9  \\
\hline
\end{tabular}
\caption{Estimate of systematic uncertainty on the initial FSW localizations at 1.9 s post-trigger\label{tab:fsw_init}.}}
\end{table}

Fits using the more complex
core-plus-tail models converged using the entire FSW Final location sample (omitting the IPN set). 
These fits and the single component fits with a Z hemisphere dependence
 for the FSW Final locations, the only single-component model
that converged with the entire sample, are reported in 
Table~\ref{tab:fsw_final}.

\begin{table}
\footnotesize{
\begin{tabular}{|l|l|l|l|c|c|c|c|c|c|l|l|}
\hline
\multicolumn{12}{|c|}{Core + Tail fits to the systematic uncertainty on the FSW Final locations}  \\
\hline
Type & Hemisphere & \multicolumn{2}{c}{Number GRB} & Core & Error & Core & Error & Tail & Error & Log$_{10}$ &  Log$_{10}$ \\
& & Point & Annuli &  $^\circ$ &  $^\circ$ &  \% & \%  & $^\circ$ & $^\circ$ & Likelihood  & Odds Factor \\
\hline
 & All-sky & 192 & 0 & 7.52 & 0.76 & 89.7 & 2.9 & 55.6 & 9.0 &  39.6 & 38.1 \\
Quadrant & $\pm$X & 192 & 0 & 8.29 & 1.34 & 85.7 & 6.2 & 59.6 & 12.9 &   &  \\
 & $\pm$Y &  &  & 7.35 & 1.03 & 92.9 & 3.7 & 62.6 & 18.5 &  40.1 & 38.5 \\
Hemi & +X & 192 & 0 & 5.99 & 1.12 & 86.3 & 4.6 & 47.2 & 9.0 &   &  \\
 & -X &  &  & 8.80 & 1.02 & 92.2 & 3.4 & 68.2 & 27.2 &  40.4 & 38.5 \\
Hemi & +Y & 192 & 0 & 7.84 & 1.16 & 92.4 & 4.9 & 35.6 & 12.3 &   &  \\
 & -Y &  &  & 6.66 & 1.17 & 84.3 & 4.8 & 62.2 & 13.2 &  41.3 & 39.5 \\
\hline
 \multicolumn{12}{|l|}{The Z-hemisphere model failed to converge. The Z-hemisphere-dependent single component is below.} \\
\hline
Single & +Z & 192 & 0 & 12.18  & 0.7 &  &  &  &  &  &   \\
 & -Z &  &  & 34.09 & 2.71 & &  &  & & 13.0  & 12.6 \\
\hline
\end{tabular}
\caption{Estimate of systematic uncertainty on the last-issued FSW localizations\label{tab:fsw_final}.}}
\end{table}

In September 2009, the {\it Fermi} rocking angle changed from $35^\circ$ to $50^\circ$.  
The rocking angle change was necessary to keep the spacecraft battery cool, but one unfortunate
effect on GBM is that more GRBs are viewed with an unfavorable detector geometry and 
the GBM localization quality may have suffered. Because of on-board hardware limitations, the
full atmospheric scattering calculation performed on the ground is replaced by a standard model table for the atmospheric
scattering contribution that assumes a zenith-pointed spacecraft.  This becomes
 increasingly unrealistic with the updated rocking 
angle but the on-board limitations do not permit multiple tables that would cover several possible {\it Fermi} pointings.   
In order to assess the effect of this change on the FSW localizations, the FSW Final location sample was split into the 43 GRBs
occurring before the change and the 149 since then.  Table~\ref{tab:fsw_rock} shows a comparison of the systematic errors on these
localizations.  Single-component fits were possible with the smaller sample and only core-plus-tail modeling was possible
with the larger sample.  It appears the worst outliers have been detected since the rocking angle change and the value of
the systematic error for the core is higher than the single model that best fits the sample localized before the
rocking angle change.  This suggests there has been a deterioration in the quality of the FSW localizations since September
2009.

\begin{table}
{\footnotesize
\begin{tabular}{|l|l|c|c|c|c|c|c|l|l|}
\hline
\multicolumn{10}{|c|}{Effect of rocking angle change on FSW localization systematic uncertainty} \\
\hline
Type & Hemisphere &  Core & Error & Core & Error & Tail & Error & Log$_{10}$ &  Log$_{10}$ \\
 &  &  $^\circ$ &  $^\circ$ &  \% & \%  & $^\circ$ & $^\circ$ & Likelihood  & Odds Factor \\
\hline
\multicolumn{10}{|c|}{FSW Final locations for 43 GRBs in initial rocking profile} \\
\hline
All-sky  & &  5.80 & 0.99 & &  &  &  &  25.3 & 24.9 \\
Hemi & +Y  &  4.52 &   1.39 & & & &  &  & \\
 & -Y & 6.98 & 1.58 &  &  &  &  & 25.6 & 25.2 \\
Quad & $\pm$X &  8.10 &   1.79 &  &  &  &  &  & \\
 & $\pm$Y & 3.82 & 1.26 &  &  &  &  & 26.2 & 25.9 \\
\hline
\multicolumn{10}{|c|}{FSW Final locations for 149 GRBs since the rocking angle change} \\
\hline
All-sky &  &  8.32 & 1.01 & 87.0 & 3.8 & 56.6 & 9.8 & 17.1 & 15.9 \\
Hemi & +X  &  7.59 &   1.32 & 84.0 & 5.7 & 48.8 & 10.1 &  & \\
 & -X & 9.14 & 1.26 & 89.9 & 4.6 & 67.7 & 27.0 & 17.3 & 15.9 \\
Hemi & +Y  &  10.21 &   1.46 & 95.1 & 5.4 & 48.2 & 29.6 &  & \\
 & -Y & 6.18 & 1.77 & 79.9 & 6.3 & 60.1 & 12.6 & 19.0 & 18.0 \\
Hemi & +Z  &  7.52 &   1.42 & 93.1 & 5.4 & 42.4 & 15.9 &  & \\
 & -Z & 10.40 & 2.32 & 67.3 & 9.9 & 63.0 & 14.8 & 21.0 & 20.1 \\
Quad & $\pm$X &  7.93 &   1.63 & 78.5 & 7.4 & 52.6 & 10.7 &  & \\
 & $\pm$Y & 9.26 & 1.09 & 93.3 & 3.8 & 71.9 & 51.7 & 17.8 & 16.8 \\
\hline
\end{tabular}
\caption{Effect of rocking angle change on the quality of FSW final locations. The systematic error
for the sample ofr 43 GRBs before the spacecraft rocking angle increased from $35^\circ$ to $50^\circ$
is modeled by a single Gaussian.  More complicated models failed to converge, possibly owing to the
small sample size.  For the 149 GRBs detected after the rocking angle change, only the core-plus-tail
models converged, probably because the systematic errors associated with the tail of the population
were too large for a single-component model to result in an acceptable fit.\label{tab:fsw_rock}}}
\end{table}

\subsection{The special case for short GRBs}
Short GRBs (SGRB, duration $<$ 2 s) are especially interesting to the follow-up community, both because they are rarer and because
their putative association with the merger of a neutron star with either a neutron star (NS-NS) or a black-hole (NS-BH) is less
well established than the connection between long GRBs and the collapse of massive stars. 
It is useful to characterize the systematic errors
of the short GRB localizations in order to assess the regions with the maximum probability of containing the source or
to calculate the probability that the GRB was in a region observed in another wavelength.    

\begin{figure}
 \includegraphics{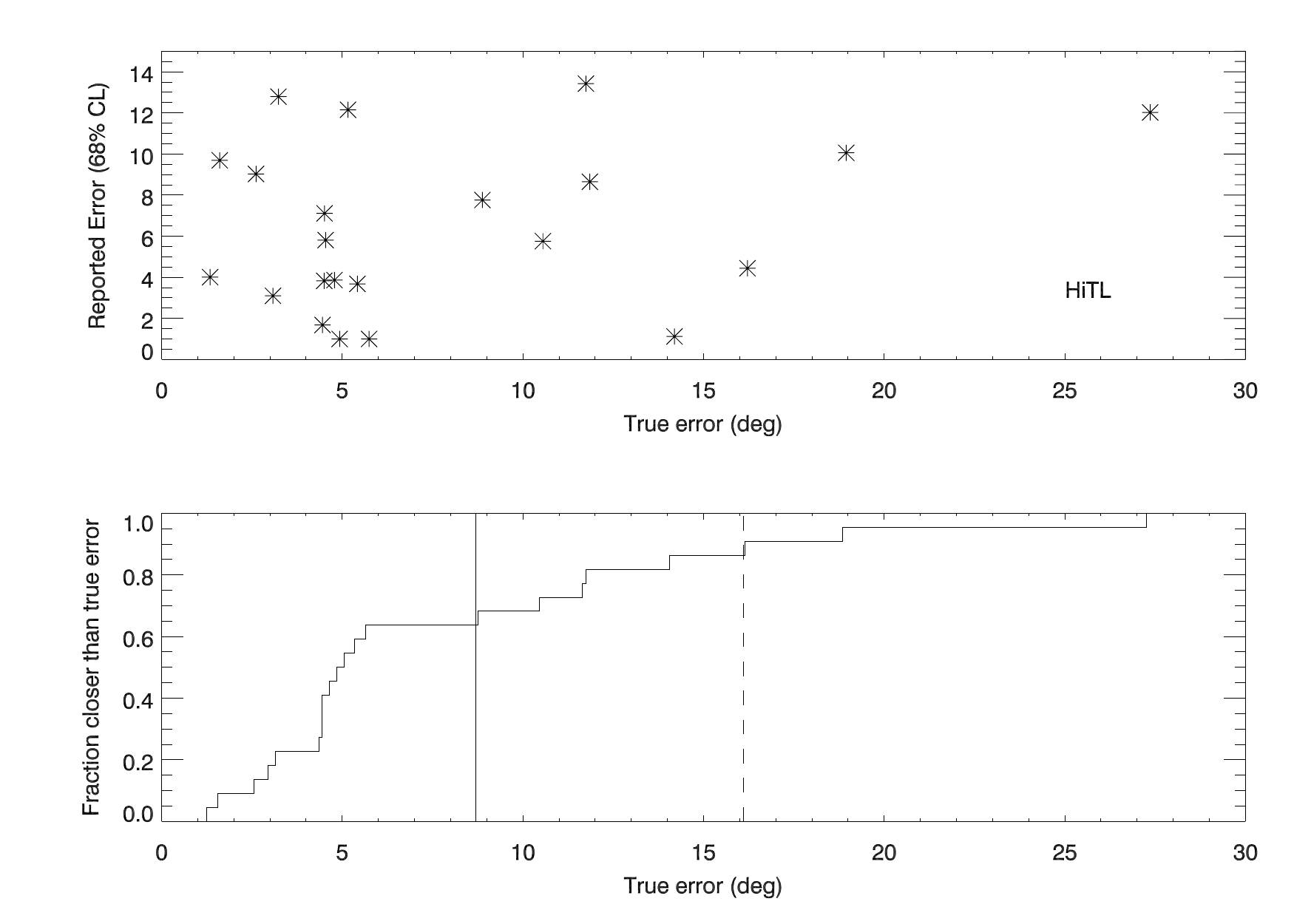}
 \caption{The reported 68\% CL uncertainty for 22 short GRBs
localized by GBM is shown in the top panel as a function of the true offset. 
The histograms show the fraction of GBM HitL localizations
lying within a given offset (degrees) from the true position.
The solid vertical line
indicates the 68\% containment radius 8.7$^\circ$ and 
the dashed vertical line the 90\%
radius of 16.1$^\circ$.
The 68\% containment region for the 68\% containment radius for the HiTL
locations lies between 5.3 and $11.7^\circ$ with the 90\% containment
radius having a lower limit of $11.6^\circ$ and being unconstrained at the
upper end.
\label{fig:locs_short_hitl}}  
 \end{figure}

\begin{figure}
 \includegraphics{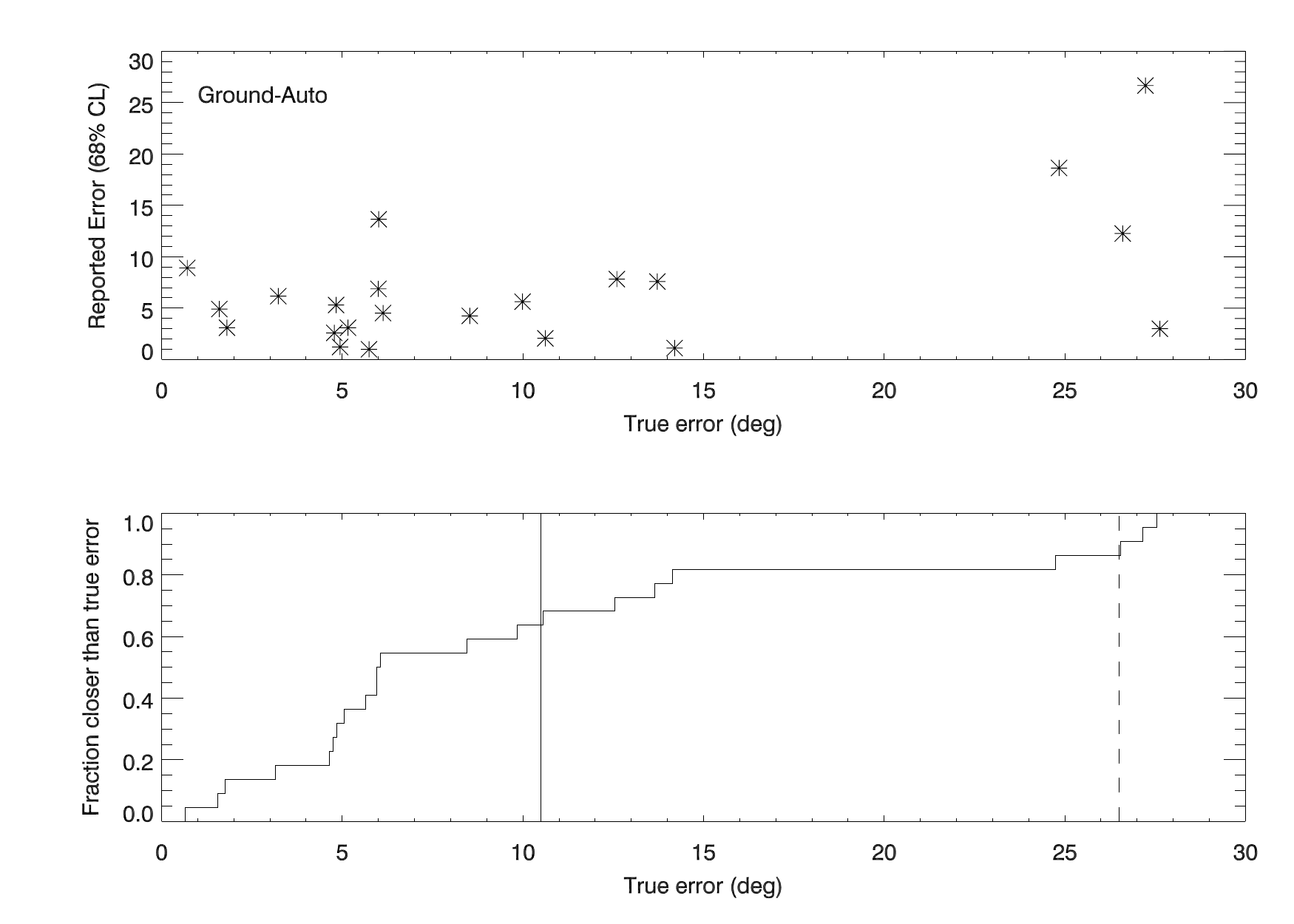}
 \caption{The reported 68\% CL uncertainty for 22 short GRBs
localized by GBM using the ground-automated process is shown in the top panel as a function of the true offset. 
The histograms show the fraction of GBM Ground-Auto localizations
lying within a given offset (degrees) from the true position.
The solid vertical line
indicates the 68\% containment radius 10.5$^\circ$ and 
the dashed vertical line the 90\%
radius of 26.5$^\circ$.
The 68\% containment region for the 68\% containment radius for the Ground-Auto
locations lies between 8.4 and $14.1^\circ$ with the 90\% containment
radius having a lower limit of $13.6^\circ$ and being unconstrained at the
upper end.
\label{fig:locs_short_ga}} 
 \end{figure}

Only 22 of the 203 reference locations in our sample are for short GRBs, and an additional 13 SGRBs have narrow IPN annuli.
The top panel of Figure~\ref{fig:locs_short_hitl}  shows the reported 68\% CL error on the GBM locations as a function of their offset from the real GRB position
for the HitL localizations of the 22 SGRBs in our sample. The cumulative distribution of
offsets from the true position appears in the bottom panel. Owing to the low fluences of SGRBs, they are typically localized
with larger statistical uncertainties than long GRBs. The 68\% and 90\% containment radii of the true positions from the GBM positions are 8.7$^\circ$ and 
16.1$^\circ$, respectively, but with large uncertainties owing to the small sample size.  
There is a steep rise in containment of the true source position at just under 6$^\circ$ from the GBM 
position after which it remains flat, though with 22 events this curve may also result from small statistics. 
Covering an area 6$^\circ$ around the reported position might be a good
strategy, regardless of the reported error on the localization, which is only weakly correlated with the
true source offset for these short events.  Ground-automated localizations of SGRBs are displayed in the same way in Figure~\ref{fig:locs_short_ga}. 
It can be seen
that although the containment radii are higher (10.5$^\circ$ and 26.5$^\circ$ for the 68\% and 90\% sample containment), 
there is a similar steep rise at just under 6$^\circ$ from the position reported by GBM so that with a limited FoV, observers would capture
50\% of the true positions using a search radius of 6$^\circ$.  

More short GRBs with known positions are required to characterize the systematic error on these events convincingly. A preliminary analysis of the
HitL 4.14g locations for SGRBs using our 
Bayesian code, which uses 13 IPN annuli in addition to the 22 reference locations, suggests a single-component systematic 
uncertainty of
$7.0 \pm 1.0^\circ$.  A core-plus-tail model with $4.7 \pm 1.0^\circ$, a core fraction of $78 \pm 2\%$ and a tail 
component of $12 \pm 5^\circ$ is mildly preferred and suggests the short burst systematic uncertainties
 are compatible with those of the GRB population as a 
whole, with the exception that the fraction in the tail is higher. This could explain the larger systematic
uncertainty
 obtained in the single-component model for the short bursts compared to the population as a whole ($5.2 \pm 0.2^\circ$).  
Attempts to characterize the systematic uncertainty
 associated with the Ground-Auto localizations fail with this small sample.   
The sample of SGRBs with IPN annuli can be increased using the large number of SGRBs seen only by GBM and Konus-Wind \citep{valentin}.

\subsection{Applying the model to the data}
Figure~\ref{fig:locs_cum_sigma} showed that only 39\% of the HitL localizations fell within the $1\sigma_{stat}$ 
statistical uncertainty region of the true reference location, $\sigma_{stat}$,
 with 70\% within $2\sigma_{stat}$.    
If we convolve the statistical uncertainty, assuming circular uncertainty regions, with the best-fit quadrant-dependent
core-plus-tail model of the systmatic uncertainty, $\sigma_{sys}$,
explored in Section 5, adding the functions in quadrature
$\sigma_{tot} = \sqrt(\sigma_{stat}^2+\sigma_{sys}^2)$, 
then we find that 67.5\% lie within the $1\sigma_{tot}$
radius and 94.5\% within $2\sigma_{tot}$.  If instead of the circular assumption we make for $\sigma_{stat}$ we 
convolve the $\chi^2$ map with the model for the systematic uncertainty that was found to be the best fit of those
we explored in Section 5, these numbers are 68.7\% and 91.4\%, respectively.  Similar results are found for the Ground-auto 
localizations.  Propagating the uncertainties on the model parameters produces a variation of 1-2\% in these containment percentages.

\section{Summary and future plans}
Using a reference sample of 203 GRBs with known locations we have found that ground-automated GBM localizations distributed 
as Ground Position GCN notices between 30 s and about a minute
after the trigger time lie within $7.6^\circ$ of the true location 68\% of the time and within $17.2^\circ$ 90\% of the time (Figure~\ref{fig:locs_cum}).  
These numbers are for the population as a whole and do not take into account the reported statistical uncertainty.  
For GRBs with
small statistical localization
 uncertainties of $1^\circ$ this improves to $2.2 - 5.6 ^\circ$ for the 68\% containment level, with the
sample of GRBs with localization errors this small insufficient to determine the 90\% containment level
(Figure~\ref{fig:locs_cum_1deg}).  
The localizations produced with human intervention (HitL) that are distributed as Final Position GCN notices 30 min to hours after the trigger are
within $5.3^\circ$ and $10.1^\circ$ of the true position for 68\% and 90\% of GRBs, respectively.  For GRBs with 
statistical localization uncertainties of $1^\circ$
the 68\% containment radius is  $3.3 - 4.7 ^\circ$ with the sample again too small 
to determine the 90\% containment radius.

An analysis of the systematic uncertainty on GBM localizations that takes into account the reported statistical 
uncertainty used, in addition to the 203
point locations, 100 IPN annuli and 9 IPN intersecting annuli that are treated as point locations, for a total of 312 
reference locations. The model that best represents the 
systematic uncertainty
for both the automated and HitL localizations includes a core component and a tail component.  For the Ground-Auto, the core
component is a Gaussian that peaks at $3.7 \pm 0.3^\circ$ and contains about 80\% of the GRB locations, with a Gaussian tail peaked at 
$14 \pm 2^\circ$.
The HitL systematic uncertainty has similar values but 90\% of the localizations are contained in the core (Table~\ref{tab:corept}).  
Both localization types show evidence for 
a dependence of the error on the GRB position in spacecraft coordinates, with bursts incident near
 the $\pm$Y-axes better localized than
those near the $\pm$X-axes (Table~\ref{tab:corept_quad}).   With three extra parameters,
the size of the reference sample is not large enough for this model to be preferred statistically.  Modifying the azimuthal area covered in these
X- and Y- axes produces a model for the HitL localizations
that is preferred over the simpler core-plus-tail model described above (Table~\ref{tab:corept_quad_slide}).  Because the parameters in this model are
well-constrained and the results are reproduced using the Ground-Auto localizations, we consider this a good model for the systematic uncertainty of 
GBM-localized GRBs.   

Applying this model to the data, using either the circular approximation to $\sigma_{stat}$ that
was input to the Bayesian model or the actual $\chi^2$-grid values to define the 
statistical confidence levels, resulted in the containment percentages of the
the reference locations within the expected total uncertainty regions.    

The quality of our localization depends on the quality of the model rates that we 
compare with our observed rates. 
This in turn relies on 
our knowledge of the detector responses both to the direct flux and to the scattered flux from both the spacecraft and the atmosphere.  Systematic effects owing to a poor choice of model
spectrum may contribute to a poor localization, as can inaccuracies in our detector response,
our mass model of the Fermi spacecraft,
and our model of scattering from the atmosphere. 
The dependence of the systematic uncertainty
 in our localization on the position of the GRB in spacecraft coordinates may offer clues to 
the major sources of these systematic errors. A systematic error that was dominated by our detector responses or by inaccuracies in our
atmospheric scattering modeling is unlikely to exhibit such a dependence on position in spacecraft coordinates, although both of these
factors may play a part.  
The $\pm$X 
sides of the spacecraft contain much of the wiring and electronics boxes, with the -X side
also housing the star trackers.  By
contrast, the $\pm$Y sides, housing the LAT radiators and solar panels, are clearer of material. 
 The quadrant dependence of $\sigma_{sys}$
implies that the bursts incident on parts of the spacecraft with fewer electronics boxes and large cable bundles have better localizations than those
arriving on a busier part of the spacecraft.  This may imply inaccuracies in our mass model lead to a miscalculation of the
observed rates that arise from scattering in the spacecraft.
We will explore the possibilities of improving this model in future work.   

The localization of short GRBs is poorer than long GRBs, with 68\% localized within $5.3 - 11.7^\circ$ of the true position 
when humans are involved and $8.4 - 14.1 ^\circ$ in the automated process.  The sample is small so these containment radii 
have large uncertainties.  An analysis of
the systematic uncertainty on the HitL locations of short bursts finds it is similar to long bursts, suggesting the difference is owed to
the poorer statistics associated with fewer counts.  Figures \ref{fig:locs_short_hitl} and \ref{fig:locs_short_ga} show that the large values
may be attributable to outliers and that 50\% of both HitL and Ground-Auto localizations of short GRBs are contained within 6$^\circ$ of the true
location.

Flight software locations suffer larger statistical and systematic uncertainties and are useful mainly to assist the LAT onboard science algorithms and to
initiate repointings of the spacecraft in response to bright GRBs (Figure~\ref{fig:fsw_loc} and Tables~\ref{tab:fsw_init} and \ref{tab:fsw_final}).  
The quality of FSW localizations has significantly declined following the
rocking profle change of {\it Fermi} in September 2009 (Table~\ref{tab:fsw_rock})
and we will endeavor to mitigate this deterioration, with enhancements subject to the limitations of onboard processing.

Improving the Ground-Auto localization to approach the quality of the HitL localization is a priority, with the first goal to replace
the peak flux localizations, which use MAXRATES packets, with fluence localizations 
using the entire count-rate time series as it arrives in real-time. This should reduce the statistical component
to the localization uncertainty on the Ground-Auto positions.
Identifying outliers in the automated process could also help, particularly for the short GRBs,
 which will not be significantly improved by the implementation
of a fluence localization.

\begin{figure}
 \includegraphics[width=18cm]{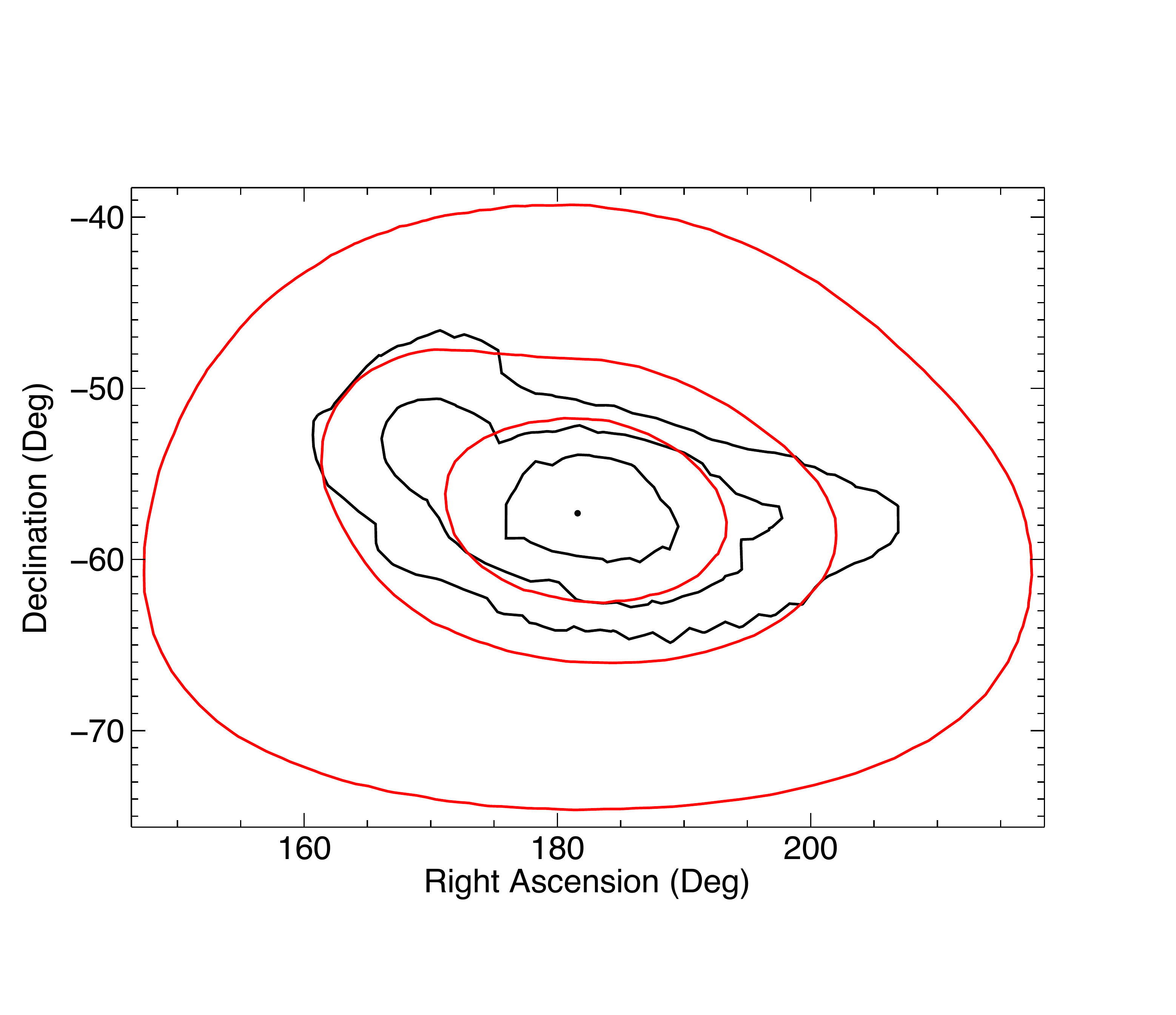}
 \caption{The $1\sigma$, $2\sigma$, $3\sigma$ contours
for GRB 080714745.  The black contours are the
statistical uncertainty $\sigma_{stat}$, returned by the localization process.
After convolving with the best-fit model for the systematic uncertainty, $\sigma_{sys}$, 
described in the text as a sum of core-plus-tail Fisher functions with parameters
that vary according to the quadrant in which the GRB is localized, the red
curves are obtained, $\sigma = \sqrt(\sigma_{stat}^2 + \sigma_{sys}^2)$.\label{fig:contour}.} 
\end{figure}

The GCN localization notices currently report a 68\% confidence-level statistical uncertainty
 that assumes a circular region around the
$\chi^2$ minimum position.  
In practice, the $\sigma_{stat}$ contours can be elliptical or irregular depending on the $\chi^2$ map returned in the minimization process.  
We have
implemented an algorithm to convolve the $\chi^2$ map with an input systematic uncertainty
 model to return $1\sigma$, $2\sigma$, and $3\sigma$ contour
maps for our localizations.  
An example is shown in Figure \ref{fig:contour}.  The contours returned by the localization process can be ragged. 
They are 
overplotted on the smoothed contours with which the systematic uncertainty, modeled as the sum of two quadrant-dependent Fisher functions
described above, has been convolved. As the systematic error is better characterized, and hopefully
 improved, the model parameters are easily modified
to return the contours that best represent the current knowledge of the localization quality.
We started delivering these contour maps to the community
in January 2014.  They are used regularly by the iPTF team in follow-up observations
that have led to the discovery of seven GRB afterglows.  The maps are available at the {\it Fermi} Science Support Center 
\footnote{http://fermi.gsfc.nasa.gov/gbm} in ascii, png, and FITS format.
Because of the real-time delivery limitations of the GCN and processing and transfer latency to the FSSC, the 
contours are available at the {\it Fermi} Science Support Center from 30 minutes to a couple of hours
after the GRB trigger.

\begin{acknowledgements}
We thank an anonymous referee for very useful contributions to this paper.
The GBM project is supported by NASA.  Support for the German contribution to GBM was provided by the Bundesministerium f\"ur 
Bildung und Forschung (BMBF) via the Deutsches
Zentrum f\"ur Luft und Raumfahrt (DLR) under contract number 50 QV 0301.
A.v.K. was supported by the Bundesministeriums f\"ur Wirtschaft und Technologie (BMWi) through DLR grant 50 OG 1101.
HFY acknowledges support by the DFG cluster of excellence ``Origin and Structure of the Universe''.
AG and SG are funded through the NASA Postdoctoral Fellowship Program. 
SF acknowledges the support of the Irish Research Council for Science, Engineering, and Technology, co-funded by Marie Curie Actions under FP7.
GF acknowledges the support of the Irish Research Council.  DT acknowledges support from Science Foundation Ireland under grant number 09-RFP-AST-2400. 
\end{acknowledgements}

\clearpage

\begin{appendix}

\section{Distributions of $\chi^2$ for GBM GRB localizations}
A grid of 41168 possible arrival directions in a spacecraft coordinate
system, $1^\circ$ apart, contains the expected count rates in each of the
12 NaI detectors for a burst originating from that direction.  
The relative count rates in the 12
detectors depend not just on the arrival direction but also on the energy spectrum of the gamma-ray source.
Minimizing $\chi^2$ on the grid of 41168 possible arrival directions based on the observed
count rates from a GRB yields the best fit arrival direction for that assumed spectral shape.  

For each of 41168 positions in the grid, {\it i}, we find 

\begin{equation}{}
\chi^2(i) =
\sum\limits_{j=1}^{12} \frac{[s(j) - b(j)- f(i)*m(j,i)]^2} {b(j)+f(i)*m(j,i)}
\end{equation} 
 
where $s(j)$ and $b(j)$ are the total observed and background rates, respectively, observed between 50 and 300~keV in detector $j$;
$m(j,i)$ are the model rates in the same energy range for detector $j$ in row $i$; and $f(i)$ is the normalization
factor for row $i$ such that

\begin{equation}{}
f(i) = \frac 
{\sum\limits_{j=1}^{12} 
[m(j,i)*(s(j) - b(j))]/s(j)} 
{\sum\limits_{j=1}^{12} 
m(j,i)^2/s(j)}
\end{equation}

We try
three different spectral models representing soft, medium, and hard GRB spectra, as described in section 2. 
The direction yielding the lowest $\chi^2$ may vary according to the spectral model, so 
we obtain up to three possible arrival directions, one for each model.
The spectral model returning the lowest $\chi^2$ is assumed to be the better fit to the data and the code therefore
selects the position yielding the lowest $\chi^2$ across all three models as the most likely arrival direction for
the GRB.

In the reference sample that includes the 203 reference locations from other satellites and the 110 
locations from the IPN, the location from the
hard spectrum model table is selected 16\% of the time, the medium 53\% and the soft 31\%.  
The medium and soft GRB spectral models produce the same burst arrival direction for 33\% of the sample and only
two GRBs are localized to the same grid point under all three spectral models.  There are no cases where
the hard and soft spectra produce the same $\chi^2$ minimum arrival direction but where
those directions are different from those
obtained under the assumption of the medium GRB spectrum.  The use of three spectral models thus produces
different optimal arrival directions for 2/3 of the GRB reference sample. The
statistical preference for the soft or medium tables over the hard table or vice versa 
typically involves tens of units of $\chi^2$ whereas the selection of soft over medium or vice versa 
typically involves fewer than 20 units of $\chi^2$.  Using any of the three models individually 
for the whole sample results in poorer localizations overall, as determined by both a larger median
distance to the true location for the whole sample and a larger value for the systematic error calculated as described
in section 5.  This suggests the assumed source spectrum does affect the quality of the localization.
We have tried different spectral models, working from the distribution of measured catalog values
for Band function parameters \citep{goldstein_1grb, dg_2grb} but have thus far
not obtained a significant improvement relative to the methods and models currently being used that are described in this paper.      
This was somewhat surprising given that the assumed spectral shapes (particularly the hard spectrum) are extreme compared to
the measured spectral shapes. We also expect that because of the scattering of high-energy photons 
off material in the spacecraft into the 50 - 300~keV energy range used for the localization process, systematic effects
arising from inaccuracies in the spacecraft mass model in our simulations result in
spectrally harder bursts being more poorly localized than bursts with a softer spectrum.   
Quantifying the effect by dividing the GRB sample, for example by hardness ratios or $E_{peak}$ values, is complicated
by the fact that short GRBs are spectrally harder, have lower fluences (and hence larger statistical uncertainties), and
are underrepresented in the GRB sample with reference locations.     
The effects of spectral modeling on source localization are still being explored and will be reported
in a future paper.  

\begin{figure}
\center{
\vbox{
 \includegraphics[width=4in]{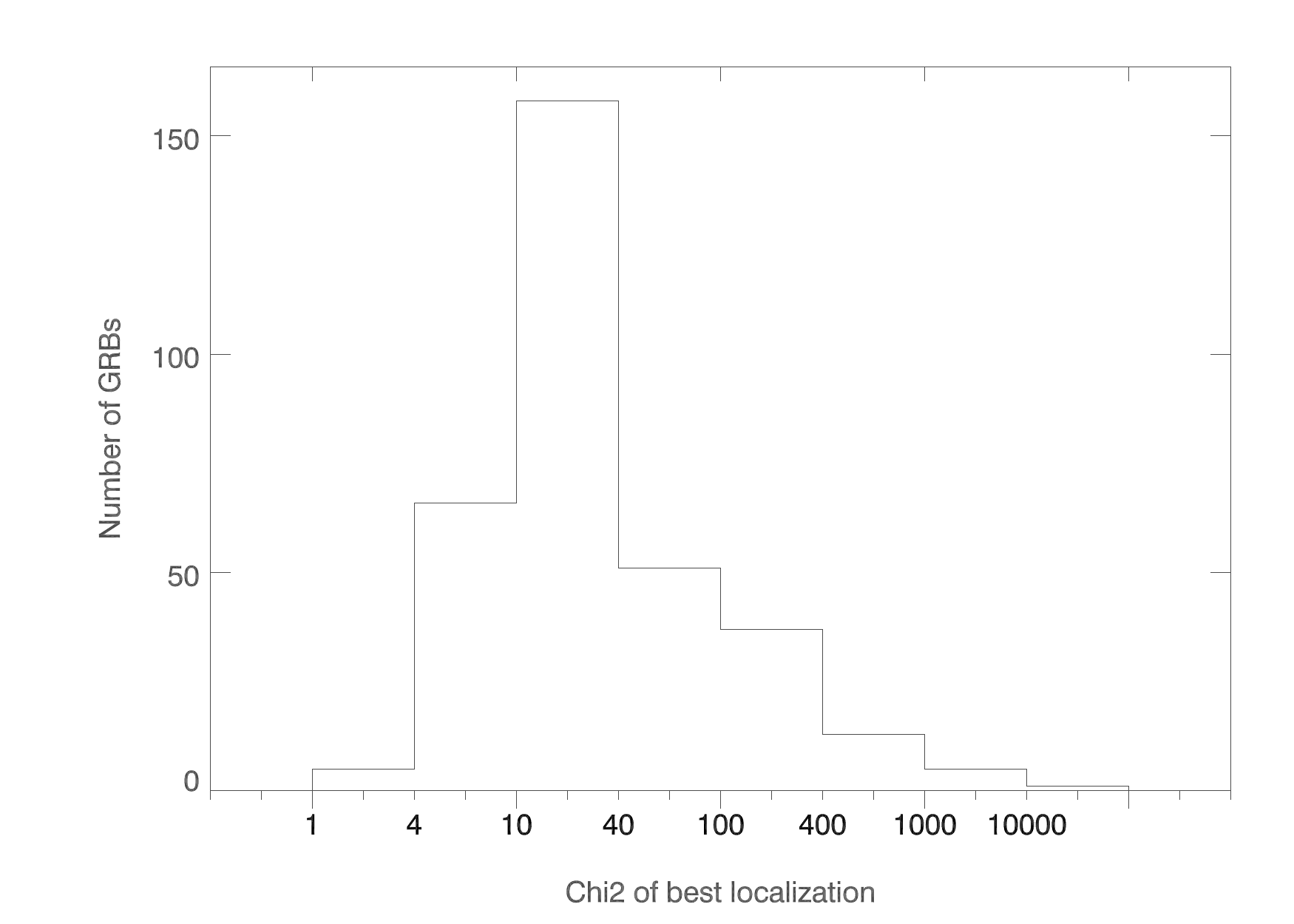}
 \includegraphics[width=4in]{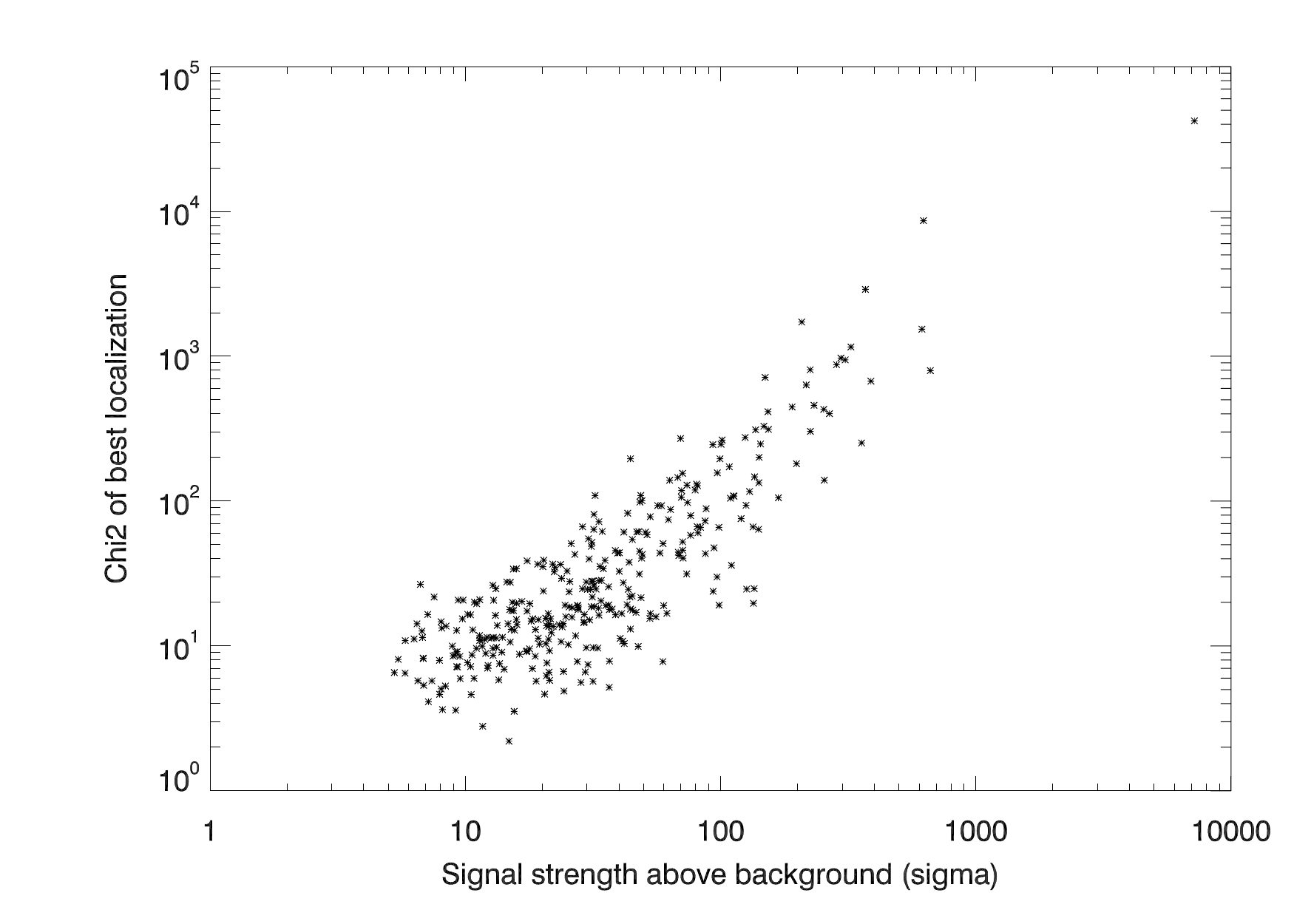}}
 \caption{
The top panel shows the distribution of minimum $\chi^2$ values for the GBM localizations of the
sample of GRBs with reference and IPN locations.  There is a strong correlation with the
intensity of the GRB, as seen in the lower panel which shows the variation of $\chi^2$ with the
strength of the data signal used in the minimization.   
\label{fig:chi2}}}
 \end{figure}

\begin{figure}
\center{
\vbox{
 \includegraphics[width=4in]{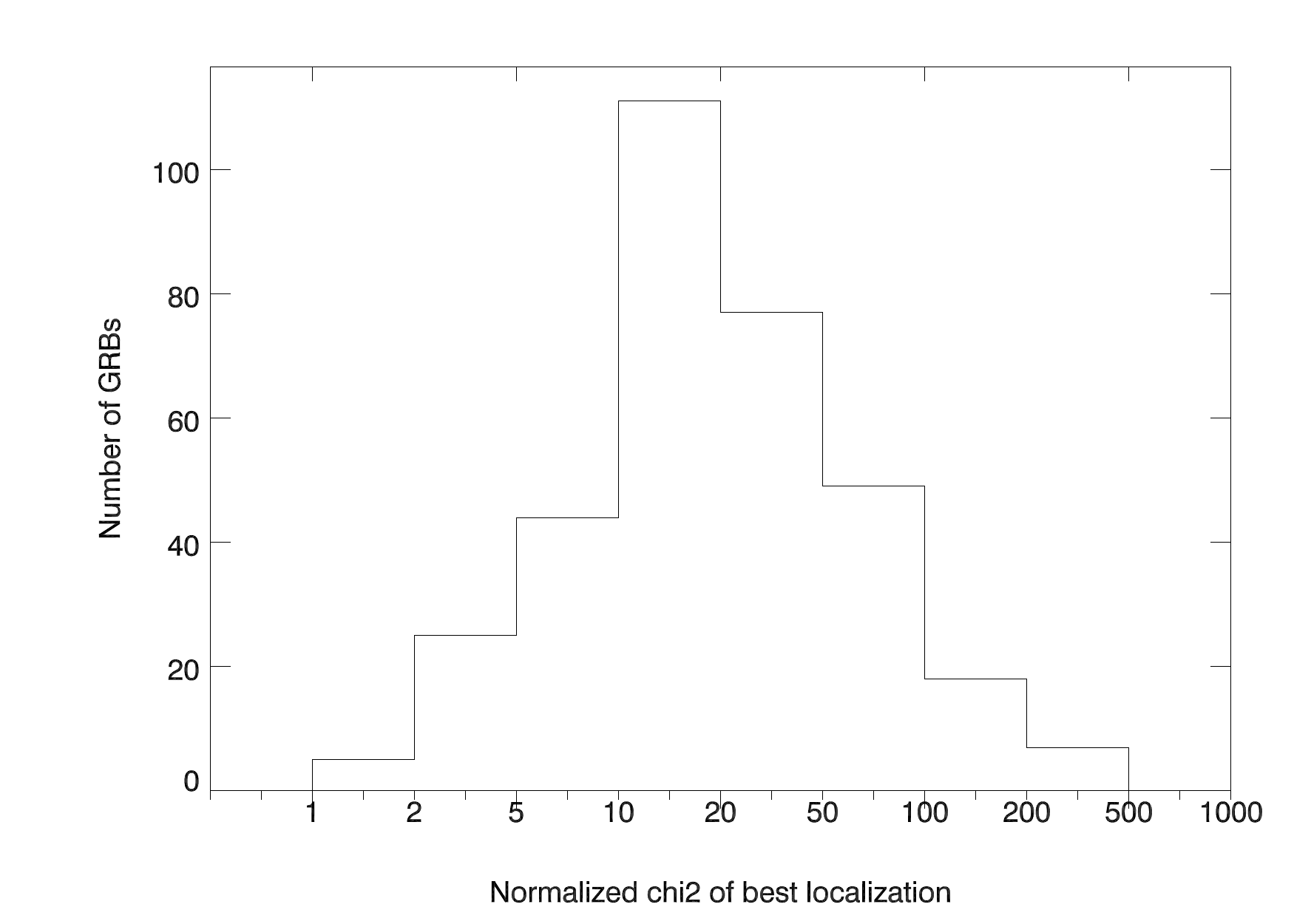}
 \includegraphics[width=4in]{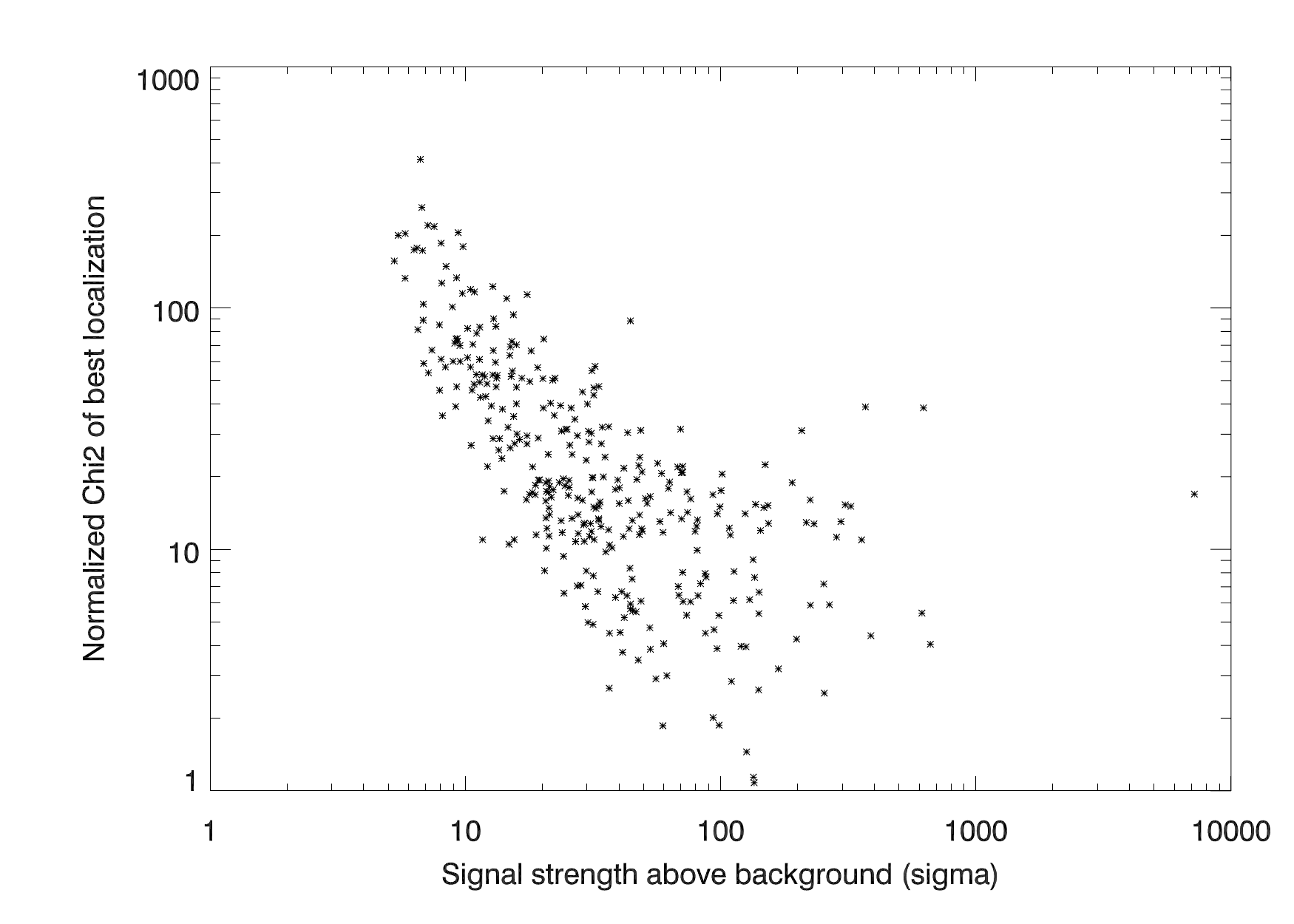}}
 \caption{
The top panel shows the distribution of minimum $\chi^2$ values for the GBM localizations of the
sample of GRBs with reference and IPN locations after normalizing the intensity
of the GRB (i.e., the observed counts that were used in the minimization) to a fiducial
burst of average intensity.  The correlation of $\chi^2$ with source signal shown in
Figure \ref{fig:chi2} is reversed (lower panel), with the weaker bursts showing higher values of $\chi^2$,
probably because poor fits to the background count rates are magnified in the
normalization process and affect the goodness-of-fit of the localization. 
\label{fig:nchi2}}}
 \end{figure}

Localization uncertainties for dim GRBs are dominated by statistics.  Bright bursts have low statistical uncertainties
and their $\chi^2$ values become larger with increasing intensity as the goodness of fit 
is affected by systematics.  Figure \ref{fig:chi2} shows the distribution of minimized $\chi^2$ values for the 
reference sample and the dependence of this value on the brightness of the GRB, represented
by the significance above the background of the data used in the localization. Figure \ref{fig:nchi2} shows the same quantities
but with the minimum $\chi^2$ calculated after normalizing the observed data rates to those of a GRB of average intensity
(1 photons/cm$^2$/s between 50 and 300 keV).  The dependence of the minimum $\chi^2$ on source intensity disappears for
bright GRBs ($> 20 \sigma$ above background) but the weaker bursts now show higher normalized $\chi^2$ values, probably because
the quality of the localization depends more on the quality of the fit to the background data than for brighter GRBs, and
systematics due to poor background fitting are magnified by the normalization of the source rates upwards to a GRB of
average intensity.  In the localization process, we use the normalized $\chi^2$ to assess 
the goodness-of-fit of the localization, rejecting localizations with normalized $\chi^2$ values above 500 as bad fits.   This threshold 
was found to be efficient in rejecting triggered events due to particle precipitation, which do not
have relative detector rates consistent with a point on the sky, without rejecting GRBs with poor localizations.   
The normalization process does not affect either the determination of the arrival direction, the 
spectral model that is selected, or the reported localization uncertainty that is calculated based on observed rather than
normalized count rates.

\newpage

\section{Reference locations}

\newpage

\begin{sidewaystable}[!h]
\caption{203 reference locations for GRBs that were also detected by GBM and which we use to assess the accuracy of GBM locations.
These locations come from {\it Swift} BAT, {\it Swift} XRT, {\it Fermi}-LAT, INTEGRAL, MAXI, Super-AGILE, and the IPN. All RA, Dec, and Err are in units of degree.\label{tab:refs}}
{\tiny
\begin{tabular}{cccccccccccccccccc}
\hline
Burst Name  & Common Name & Trigger Time & \multicolumn{3}{c}{FSW$_\text{first}$ for Ref.} & \multicolumn{3}{c}{FSW$_\text{last}$ for Ref.} & \multicolumn{3}{c}{Ground Auto for Ref.} & \multicolumn{3}{c}{H.i.t.L for Ref.} & \multicolumn{2}{c}{Ref. Location} &  Reference \\
 & & (UT) & (RA & Dec & Err) & (RA & Dec & Err) & (RA & Dec & Err) & (RA & Dec & Err) & (RA & Dec) &  \\ 
\hline
080714745 & GRB 080714A & 17:52:54.03 & - & - & - & - & - & - & 185.3 & -57.7 & 3.8 & 183.5 & -57.5 & 3.9 &  188.1 & -60.2 &  Barthelmy et al., GCN 7979 \\   
080723557 & GRB 080723B & 13:22:21.38 & - & - & - & - & - & - & 177.6 & -58.8 & 1.2 & 175.1 & -60.7 & 1.0 &  176.8 & -60.2 &  Gotz et al., GCN 8002 \\
080725435 & GRB 080725A & 10:26:09.06 & - & - & - & - & - & - & 120.2 & -13.4 & 4.1 & 123.1 & -23.1 & 2.2 &  121.7 & -14.0 &  Krimm et al., GCN 8014 \\
080727964 & GRB 080727C & 23:07:46.22 & - & - & - & - & - & - & 43.5 & 59.8 & 4.2 & 40.0 & 62.2 & 2.7 &  32.6 & 64.1 &  Fenimore et al., GCN 8044 \\
080804972 & GRB 080804A & 23:20:14.88 & - & - & - & - & - & - & 324.1 & -52.7 & 3.9 & 320.8 & -52.7 & 2.4 &  328.7 & -53.2 &  Markwardt et al., GCN 8067 \\
080810549 & GRB 080810A & 13:10:12.59 & - & - & - & - & - & - & 358.3 & -5.7 & 5.0 & 355.8 & 5.5 & 2.2 &  356.8 & 0.3 &  Sakamoto et al., GCN 8082 \\
\hline
\end{tabular}
}
\end{sidewaystable}

\begin{sidewaystable}[!h]
{\tiny
\begin{tabular}{cccccccccccccccccc}
\hline
Burst Name  & Common Name & Trigger Time & \multicolumn{3}{c}{FSW$_\text{first}$ for Ref.} & \multicolumn{3}{c}{FSW$_\text{last}$ for Ref.} & \multicolumn{3}{c}{Ground Auto for Ref.} & \multicolumn{3}{c}{H.i.t.L for Ref.} & \multicolumn{2}{c}{Ref. Location} &  Reference \\
 & & (UT) & (RA & Dec & Err) & (RA & Dec & Err) & (RA & Dec & Err) & (RA & Dec & Err) & (RA & Dec) &  \\ 
\hline
090328401 & GRB 090328A & 09:36:46.52 & 79.3 & -46.7 & 26.3 & 80.0 & -45.6 & 5.3 & 95.0 & -44.5 & 2.1 & 95.9 & -45.1 & 1.0 &  90.9 & -42.0 &  McEnery et al., GCN 9044 \\
090422150 & GRB 090422A & 03:35:17.07 & 296.9 & 36.1 & 18.4 & 296.9 & 36.1 & 18.4 & 297.7 & 42.1 & 8.5 & 296.3 & 33.2 & 10.3 &  294.7 & 40.4 &  Markwardt et al., GCN 9195 \\
090423330 & GRB 090423A & 07:55:25.40 & 151.3 & 11.0 & 31.9 & 151.3 & 11.0 & 31.9 & 144.5 & 9.1 & 50.0 & 154.9 & 13.5 & 11.2 &  148.9 & 18.2 &  Palmer et al., GCN 9204 \\       
090424592 & GRB 090424A & 14:12:08.67 & 177.4 & 8.0 & 3.5 & 177.3 & 7.8 & 3.6 & 191.4 & 18.0 & 1.0 & 191.5 & 18.1 & 1.6 &  189.5 & 16.8 &  Sakamoto et al., GCN 9231 \\
090509215 & GRB 090509A & 05:10:05.72 & 220.2 & -36.7 & 26.0 & 241.9 & -28.8 & 15.5 & 237.5 & -27.7 & 10.1 & 230.1 & -27.7 & 5.7 &  241.4 & -28.4 &  Tueller et al., GCN 9335 \\
090510016 & GRB 090510A & 00:22:59.98 & 340.8 & -33.3 & 4.0 & 340.8 & -33.3 & 4.0 & 335.1 & -31.3 & 1.3 & 335.0 & -31.3 & 1.0 &  333.6 & -26.6 &  Ukwatta et al., GCN 9337 \\
\hline
\end{tabular}
}
\end{sidewaystable}

\begin{sidewaystable}[!h]
{\tiny
\begin{tabular}{cccccccccccccccccc}
\hline
Burst Name  & Common Name & Trigger Time & \multicolumn{3}{c}{FSW$_\text{first}$ for Ref.} & \multicolumn{3}{c}{FSW$_\text{last}$ for Ref.} & \multicolumn{3}{c}{Ground Auto for Ref.} & \multicolumn{3}{c}{H.i.t.L for Ref.} & \multicolumn{2}{c}{Ref. Location} &  Reference \\
 & & (UT) & (RA & Dec & Err) & (RA & Dec & Err) & (RA & Dec & Err) & (RA & Dec & Err) & (RA & Dec) &  \\ 
\hline
100116897 & GRB 100116A & 21:31:00.25 & 300.0 & 12.3 & 13.5 & 300.0 & 12.3 & 13.5 & 307.0 & 26.0 & 3.1 & 308.4 & 22.7 & 1.2 &  305.0 & 14.5 &  McEnery et al., GCN 10333 \\
100117879 & GRB 100117A & 21:06:19.67 & 6.1 & -6.2 & 11.9 & 6.1 & -6.2 & 11.9 & 9.9 & 8.3 & 5.7 & 11.2 & 9.0 & 5.8 &  11.3 & -1.6 &  Markwardt et al., GCN 10338 \\
100206563 & GRB 100206A & 13:30:05.40 & 44.7 & 17.6 & 11.5 & 44.7 & 17.6 & 11.5 & 74.6 & 22.7 & 3.0 & 63.9 & 13.9 & 4.5 &  47.2 & 13.2 &  Sakamoto et al., GCN 10379 \\
100212588 & GRB 100212A & 14:07:22.30 & 7.3 & 48.7 & 10.6 & 7.3 & 48.7 & 10.6 & 4.7 & 45.5 & 4.8 & 2.9 & 44.8 & 4.3 &  356.4 & 49.5 &  Ukwatta et al., GCN 10404 \\    
100216422 & GRB 100216A & 10:07:00.19 & 148.9 & 46.4 & 17.0 & 148.9 & 46.4 & 17.0 & 157.7 & 49.0 & 7.6 & 155.5 & 47.4 & 8.7 &  154.3 & 35.5 & Cummings et al., GCN 10428 \\
100225115 & GRB 100225A & 02:45:31.15 & 310.2 & -60.7 & 21.3 & 302.3 & -67.9 & 10.4 & 319.8 & -56.3 & 4.5 & 312.4 & -54.7 & 3.3 &  310.3 & -59.4 &  Piron et al., GCN 10450 \\
\hline
\end{tabular}
}
\end{sidewaystable}

\begin{sidewaystable}[!h]
{\tiny
\begin{tabular}{cccccccccccccccccc}
\hline
Burst Name  & Common Name & Trigger Time & \multicolumn{3}{c}{FSW$_\text{first}$ for Ref.} & \multicolumn{3}{c}{FSW$_\text{last}$ for Ref.} & \multicolumn{3}{c}{Ground Auto for Ref.} & \multicolumn{3}{c}{H.i.t.L for Ref.} & \multicolumn{2}{c}{Ref. Location} &  Reference \\
 & & (UT) & (RA & Dec & Err) & (RA & Dec & Err) & (RA & Dec & Err) & (RA & Dec & Err) & (RA & Dec) &  \\ 
\hline
110112934 & GRB 110112B & 22:24:55.30 & 21.6 & 44.5 & 9.7 & 21.6 & 44.5 & 9.7 & 25.8 & 42.8 & 5.8 & 25.9 & 44.0 & 4.7 &  10.6 & 64.4 &  Mereghetti et al., GCN 11562 \\
110119931 & GRB 110119A & 22:21:00.17 & - & - & - & 359.6 & -1.6 & 9.8 & 349.6 & 6.0 & 4.4 & 345.4 & 8.0 & 2.5 &  348.6 & 6.0 &  Baumgartner et al., GCN 11584 \\
110120666 & GRB 110120A & 15:59:39.23 & 65.1 & -22.6 & 5.3 & 65.6 & -23.0 & 4.9 & 65.7 & -18.3 & 1.0 & 64.8 & -16.8 & 1.1 &  61.6 & -12.0 &  Omodei et al., GCN 11597 \\
110128073 & GRB 110128A & 01:44:36.44 & 207.5 & 7.3 & 27.5 & 207.5 & 7.3 & 27.5 & 202.7 & 22.3 & 14.0 & 204.4 & 19.7 & 11.4 &  193.9 & 28.1 &  Cummings et al., GCN 11614 \\
110201399 & GRB 110201A & 09:35:10.26 & 177.3 & 59.8 & 11.3 & 177.3 & 59.8 & 11.3 & 164.8 & 84.4 & 6.5 & 166.3 & 87.4 & 4.6 &  137.6 & 88.6 &  Krimm et al., GCN 11624 \\
110207470 & GRB 110207A & 11:17:20.29 & 14.3 & -12.3 & 13.4 & 14.3 & -12.3 & 13.4 & 359.5 & 4.7 & 13.4 & 15.7 & -10.0 & 5.5 &  12.5 & -10.8 &  Palmer et al., GCN 11664 \\
\hline
\end{tabular}
}
\end{sidewaystable}

\begin{sidewaystable}[!h]
{\tiny
\begin{tabular}{cccccccccccccccccc}
\hline
Burst Name  & Common Name & Trigger Time & \multicolumn{3}{c}{FSW$_\text{first}$ for Ref.} & \multicolumn{3}{c}{FSW$_\text{last}$ for Ref.} & \multicolumn{3}{c}{Ground Auto for Ref.} & \multicolumn{3}{c}{H.i.t.L for Ref.} & \multicolumn{2}{c}{Ref. Location} &  Reference \\
 & & (UT) & (RA & Dec & Err) & (RA & Dec & Err) & (RA & Dec & Err) & (RA & Dec & Err) & (RA & Dec) &  \\ 
\hline
120712571 & GRB 120712A & 13:42:25.61 & 175.4 & -23.2 & 11.8 & 165.7 & -28.0 & 8.6 & 172.0 & -20.8 & 3.7 & 175.2 & -19.8 & 1.9 &  169.6 & -20.1 & Barthelmy et al., GCN 13455 \\
120728934 & GRB 120728A & 22:25:12.74 & 167.1 & -45.7 & 21.0 & 167.1 & -45.7 & 21.0 & 131.5 & -57.0 & 6.8 & 138.2 & -54.0 & 6.2 &  137.1 & -54.4 & Cummings et al., GCN 13525 \\
120729456 & GRB 120729A & 10:56:12.67 & - & - & - & 16.5 & 51.5 & 10.8 & 17.5 & 60.2 & 2.4 & 17.0 & 49.2 & 3.0 &  13.1 & 49.9 & Ukwatta et al., GCN 13530 \\
120811649 & GRB 120811C & 15:34:55.09 & 215.3 & 56.6 & 20.4 & 215.3 & 56.4 & 17.2 & 225.9 & 50.1 & 10.8 & 214.4 & 55.8 & 5.3 &  199.7 & 62.3 & Barlow et al., GCN 13622 \\
120817168 & GRB 120817B & 04:02:29.72 & - & - & - & 233.5 & -38.2 & 3.3 & 12.5 & -21.6 & 4.6 & 12.3 & -23.8 & 3.9 &  8.3 & -26.5 & Golenetskii et al., GCN 13670 \\
120907017 & GRB 120907A & 00:24:24.51 & 80.9 & 0.6 & 14.6 & 80.9 & 0.6 & 14.6 & 77.3 & -2.5 & 12.0 & 74.1 & -8.5 & 11.7 &  74.7 & -9.3 & Racusin et al., GCN 13716 \\
\hline
\end{tabular}
}
\end{sidewaystable}

\begin{sidewaystable}[!h]
\caption{134 locations for GRBs that were also triangulated by IPN. All RA, Dec, Err, and Width are in units of degree. $^1$These 9 bursts have 2 intersecting arcs, in which the columns are as follows: RA, Dec, box side 1, box side 2.\label{tab:ipns}}
{\tiny
\begin{tabular}{ccccccccccccccccccc}
\hline
Burst Name  & Common Name & Trigger Time & \multicolumn{4}{c}{IPN Annulus} & \multicolumn{3}{c}{FSW$_\text{first}$ for IPN} & \multicolumn{3}{c}{FSW$_\text{last}$ for IPN} & \multicolumn{3}{c}{Ground Auto for IPN} & \multicolumn{3}{c}{H.i.t.L. for IPN} \\
 & & (UT) & (RA & Dec & Err & Width)$^1$ & (RA & Dec & Err) & (RA & Dec & Err) & (RA & Dec & Err) & (RA & Dec & Err) \\ 
\hline
080715950 & GRB 080715A & 22:48:40.17 & 148.6 & 14.5 & 60.5 & 0.3 & - & - & - & - & - & - & 210.6 & 12.8 & 3.3 & 213.3 & 12.2 & 2.5\\
080723985 & GRB 080723D & 23:37:42.71 & 158.4 & 10.2 & 70.9 & 0.1 & - & - & - & - & - & - & 57.7 & 78.7 & 1.3 & 52.9 & 79.1 & 1.0\\
080724401 & GRB 080724A & 09:37:40.61 & - & - & - & - & - & - & - & - & - & - & 356.2 & 32.6 & 1.6 & 357.1 & 32.0 & 1.1\\
080730520 & GRB 080730A & 12:29:15.41 & - & - & - & - & - & - & - & - & - & - & 246.3 & 5.5 & 4.1 & 246.3 & 5.5 & 2.5\\
080730786 & GRB 080730B & 18:51:38.19 & 165.9 & 6.5 & 79.0 & 0.1 & - & - & - & - & - & - & 246.6 & 27.7 & 2.0 & 246.6 & 27.7 & 1.9\\
080802386 & GRB 080802A & 09:15:10.53 & - & - & - & - & - & - & - & - & - & - & 185.9 & 22.0 & 7.9 & 157.5 & 40.4 & 14.2\\
\hline
\end{tabular}
}
\end{sidewaystable}

\begin{sidewaystable}[!h]
{\tiny
\begin{tabular}{ccccccccccccccccccc}
\hline
Burst Name  & Common Name & Trigger Time & \multicolumn{4}{c}{IPN Annulus} & \multicolumn{3}{c}{FSW$_\text{first}$ for IPN} & \multicolumn{3}{c}{FSW$_\text{last}$ for IPN} & \multicolumn{3}{c}{Ground Auto for IPN} & \multicolumn{3}{c}{H.i.t.L. for IPN} \\
 & & (UT) & (RA & Dec & Err & Width) & (RA & Dec & Err) & (RA & Dec & Err) & (RA & Dec & Err) & (RA & Dec & Err) \\ 
\hline
090108020 & GRB 090108A & 00:29:02.37 & 268.1 & -20.2 & 70.7 & 0.5 & 258.6 & 53.9 & 5.5 & 258.6 & 53.9 & 5.5 & 259.8 & 44.9 & 4.2 & 257.4 & 44.0 & 4.0\\
090112729 & GRB 090112B & 17:30:15.46 & 103.5 & 23.7 & 86.5 & 0.1 & 194.1 & 19.3 & 8.4 & 194.1 & 19.3 & 8.4 & 193.8 & 24.9 & 2.1 & 193.3 & 25.9 & 1.2\\
090131090$^1$ & GRB 090131A & 02:09:21.15 & 349.21$^1$ & 18.47$^1$ & 0.0245$^1$ & 0.0612$^1$ & 33.4 & 28.1 & 29.4 & 353.3 & 16.1 & 3.9 & 352.9 & 20.5 & 1.6 & 352.4 & 20.9 & 1.2\\
090202347 & GRB 090202A & 08:19:30.41 & 332.2 & -12.3 & 56.9 & 0.4 & 260.3 & -15.9 & 22.4 & 260.4 & -15.9 & 7.4 & 280.7 & -9.6 & 4.7 & 276.4 & -4.8 & 3.3\\
090207777 & GRB 090207A & 18:39:10.84 & 158.7 & 8.7 & 78.3 & 0.2 & 244.5 & 14.7 & 9.8 & 243.5 & 34.9 & 9.8 & 242.3 & 38.0 & 4.2 & 243.5 & 36.9 & 4.2\\
090227310 & GRB 090227A & 07:25:57.01 & 320.3 & -16.6 & 46.8 & 0.2 & 9.1 & -42.5 & 9.2 & 4.2 & -45.9 & 7.3 & 2.5 & -42.2 & 2.2 & 2.5 & -42.1 & 1.2\\
\hline
\end{tabular}
}
\end{sidewaystable}

\begin{sidewaystable}[!h]
{\tiny
\begin{tabular}{ccccccccccccccccccc}
\hline
Burst Name  & Common Name & Trigger Time & \multicolumn{4}{c}{IPN Annulus} & \multicolumn{3}{c}{FSW$_\text{first}$ for IPN} & \multicolumn{3}{c}{FSW$_\text{last}$ for IPN} & \multicolumn{3}{c}{Ground Auto for IPN} & \multicolumn{3}{c}{H.i.t.L. for IPN} \\
 & & (UT) & (RA & Dec & Err & Width) & (RA & Dec & Err) & (RA & Dec & Err) & (RA & Dec & Err) & (RA & Dec & Err) \\ 
\hline
090718762 & GRB 090718B & 18:17:42.85 & 242.5 & -20.5 & 32.5 & 0.1 & 267.5 & -43.5 & 16.5 & 273.5 & -37.3 & 4.3 & 271.9 & -35.9 & 1.6 & 273.9 & -35.0 & 1.1\\
090719063 & GRB 090719A & 01:31:26.62 & 242.7 & -20.5 & 74.6 & 0.1 & 200.8 & -41.0 & 4.3 & 0.6 & -52.3 & 3.8 & 340.4 & -70.0 & 1.0 & 338.1 & -69.4 & 1.0\\
090720710 & GRB 090720B & 17:02:56.91 & 274.5 & -23.9 & 61.1 & 0.7 & 204.5 & -55.4 & 4.5 & 204.5 & -55.4 & 4.5 & 203.7 & -57.5 & 1.8 & 224.2 & -56.2 & 4.7\\
090802235 & GRB 090802A & 05:39:03.09 & - & - & - & - & 84.1 & 30.9 & 5.1 & 84.1 & 30.9 & 5.1 & 84.3 & 34.1 & 3.6 & 84.3 & 34.1 & 4.2\\
090804940 & GRB 090804A & 22:33:20.02 & 170.6 & 2.4 & 44.3 & 0.1 & 127.2 & -9.4 & 4.3 & 129.5 & -14.0 & 4.2 & 129.5 & -11.8 & 1.0 & 128.6 & -12.2 & 1.0\\
090809978 & GRB 090809B & 23:28:14.62 & 175.0 & -0.2 & 76.2 & 0.3 & 87.9 & 8.8 & 9.3 & 88.0 & 8.7 & 4.2 & 97.1 & 0.1 & 1.1 & 97.7 & -0.7 & 1.0\\
\hline
\end{tabular}
}
\end{sidewaystable}

\begin{sidewaystable}[!h]
{\tiny
\begin{tabular}{ccccccccccccccccccc}
\hline
Burst Name  & Common Name & Trigger Time & \multicolumn{4}{c}{IPN Annulus} & \multicolumn{3}{c}{FSW$_\text{first}$ for IPN} & \multicolumn{3}{c}{FSW$_\text{last}$ for IPN} & \multicolumn{3}{c}{Ground Auto for IPN} & \multicolumn{3}{c}{H.i.t.L. for IPN} \\
 & & (UT) & (RA & Dec & Err & Width) & (RA & Dec & Err) & (RA & Dec & Err) & (RA & Dec & Err) & (RA & Dec & Err) \\ 
\hline
100131730 & GRB 100131A & 17:30:57.68 & 132.4 & 22.5 & 16.4 & 0.3 & 117.0 & 26.2 & 3.9 & 117.0 & 26.2 & 3.9 & 131.2 & 24.8 & 1.3 & 117.5 & 13.8 & 1.5\\
100205490 & GRB 100205B & 11:45:38.26 & 115.9 & 18.2 & 41.0 & 0.6 & 126.6 & -33.4 & 21.3 & 134.7 & -20.0 & 19.2 & 133.3 & -21.4 & 8.6 & 132.6 & -22.2 & 7.1\\
100211440 & GRB 100211A & 10:33:35.17 & - & - & - & - & 120.9 & 31.8 & 5.4 & 120.9 & 31.8 & 5.4 & 138.4 & 30.2 & 3.8 & 133.3 & 29.0 & 2.3\\
100212550 & GRB 100212B & 13:11:45.47 & 117.1 & 19.3 & 21.6 & 0.3 & 138.7 & 33.9 & 7.3 & 131.4 & 34.1 & 6.6 & 136.0 & 33.8 & 2.0 & 134.3 & 32.3 & 1.9\\
100223110 & GRB 100223A & 02:38:09.31 & 125.4 & 23.8 & 25.0 & 0.1 & 105.9 & -7.5 & 5.3 & 105.9 & -7.5 & 5.3 & 94.2 & 3.0 & 6.0 & 102.1 & 5.6 & 7.7\\
100224112 & GRB 100224B & 02:40:55.48 & 306.0 & -19.3 & 38.2 & 0.2 & 269.8 & -28.6 & 15.3 & 266.2 & -9.5 & 6.2 & 271.2 & -20.8 & 1.9 & 270.0 & -18.7 & 2.2\\
\hline
\end{tabular}
}
\end{sidewaystable}

\FloatBarrier

\newpage

\section{Dependence of the single-component systematic error on GRB location in spacecraft coordinate system}
Results for single-component models exploring the effects of GRB location  
in the different spacecraft
hemispheres and quadrants are shown in Table~\ref{tab:single_geom}. The first entry in the table
repeats the values from Table~\ref{tab:single} showing the single Gaussian fits for each localization type.
The parameter values and odds ratio for the simplest single component fit can be compared with those
for the models that split the GRBs according to position in spacecraft coordinates.
GRBs localized with HitL 4.13 in the +Z  and +Y hemispheres were
found to have a lower systematic error than those in the -Z and -Y hemispheres and separating
the sample into hemispheres was slightly preferred over a single component for the whole sky. 
GRBs in the $\pm$Y quadrants
appear to have smaller systematic errors than those in the $\pm$X quadrants, and splitting the GRBs
into quadrants is slightly preferred over the all-sky single-component model, but   
the odds ratio for this model was, however, still much
lower than the core-plus-tail model in Table~\ref{tab:corept}. which is thus greatly preferred over
any of the single-component models. 
The odds ratio for these models was, however, still much
lower than the core-plus-tail model in Table~\ref{tab:corept} . 

\begin{table}
{\footnotesize
\begin{tabular}{|l|l|l|l|c|c|l|l|}
\hline
\multicolumn{8}{|c|}{Single-Component Gaussian with hemisphere dependence} \\
\hline
Type & Hemisphere & \multicolumn{2}{|c|}{Number GRB} & Peak & Error & Log$_{10}$ & Log$_{10}$ \\
& & Point & Annuli & $^\circ$ & $^\circ$ & Likelihood  & Odds Factor \\
\hline
Ground-Auto & All-sky & 208 & 100 & 6.62 & 0.29 & 155.2 & 154.3 \\
Ground-Auto & +X & 208 & 100 & 6.95 & 0.42 &  &  \\
 & -X &  &  & 6.25 & 0.41 & 155.5 & 154.0 \\
Ground-Auto & +Y & 208 & 100 & 7.02 & 0.44 &  &  \\
 & -Y &  &  & 6.21 & 0.39 & 155.7 & 154.2 \\
Ground-Auto & +Z & 208 & 100 & 6.48 & 0.33 &  &  \\
 & -Z &  &  & 7.10 & 0.64 & 155.4 & 154.0 \\
Ground-Auto & $\pm$X & 208 & 100 & 7.27 & 0.45 &  &  \\
 & $\pm$Y &  &  & 5.96 & 0.38 & 156.3 & 154.8 \\
\hline
HitL 4.14g & All-sky & 212 & 100 & 5.15 & 0.22 & 224.6 & 223.6 \\
HitL 4.14g & +X & 212 & 100 & 5.19 & 0.31 &  &  \\
 & -X &  &  & 5.06 & 0.32 & 224.6 & 222.9 \\
HitL 4.14g & +Y & 212 & 100 & 5.26 & 0.32 &  &  \\
 & -Y &  &  & 5.02 & 0.31 & 224.7 & 222.9 \\
HitL 4.14g & +Z & 212 & 100 & 4.98 & 0.24 &  &  \\
 & -Z &  &  & 5.68 & 0.53 & 225.0 & 223.3 \\
HitL 4.14g & $\pm$X & 212 & 100 & 5.41 & 0.35 &  &  \\
 & $\pm$Y &  &  & 4.93 & 0.29 & 224.9 & 223.2 \\
\hline
HitL 4.13 & All-sky & 211 & 100 & 6.23 & 0.26 & 192.6 & 191.7 \\
HitL 4.13 & +X & 211 & 100 & 6.04 & 0.35 &  &  \\
 & -X &  &  & 6.41 & 0.37 & 192.7 & 191.1 \\
HitL 4.13 & +Y & 211 & 100 & 6.84 & 0.39 &  &  \\
 & -Y &  &  & 5.57 & 0.34 & 194.0 & 192.4 \\
HitL 4.13 & +Z & 211 & 100 & 5.90 & 0.27 &  &  \\
 & -Z &  &  & 7.57 & 0.66 & 194.0 & 192.5 \\
HitL 4.13 & $\pm$X & 211 & 100 & 6.74 & 0.39 &  &  \\
 & $\pm$Y &  &  & 5.75 & 0.33 & 193.5 & 191.8 \\
\hline
\end{tabular}
\caption{Geometry-dependent single-component fits to systematic uncertainties on GBM localizations.\label{tab:single_geom}}}
\end{table}

\section{Dependence of the core-plus-tail systematic error model on GRB spacecraft hemisphere}

Results for core-plus-tail models that explore the effect of GRB position in different 
spacecraft hemispheres are shown in 
Table~\ref{tab:corept_hemi}.  The odds ratios for these hemisphere-dependent models should
be compared with the simplest all-sky core-plus-tail, repeated from Table~\ref{tab:corept} as the first
entry.
Only HitL 4.14g showed a difference from the all-sky core-plus-tail model,
with lower parameter values in the +Z hemisphere.
By comparing the odds ratios for the two models, we can see 
that this hemisphere-dependent model is not preferred statistically,
but the parameter values are different enough that with a much larger reference sample,
this model may become favored over a single core-plus-tail model.  HitL 4.13 showed
no parameter difference for the core-plus-tail division into Z hemispheres, 
and the Ground-Auto localization sample failed to converge using this 6-parameter model. 
In general the division into 
hemispheres produced similar parameters with larger uncertainties; some juggling of events between
the core and the tail can account for the slight differences in parameter values, with the exception of HitL 4.14g
and the Z hemisphere division that appears promising when a larger sample becomes available.  

\begin{table}
{\footnotesize
\begin{tabular}{|l|l|l|l|c|c|c|c|c|c|l|l|}
\hline
\multicolumn{12}{|c|}{Core + Tail (2 Gaussians) with hemisphere dependence}  \\
\hline
Type & Hemisphere & \multicolumn{2}{c}{Number GRB} & Core & Error & Core & Error & Tail & Error & Log$_{10}$ &  Log$_{10}$ \\
& & Point & Annuli &  $^\circ$ &  $^\circ$ &  \% & \%  & $^\circ$ & $^\circ$ & Likelihood  & odds Factor \\
\hline
Ground-Auto & All-sky & 208 & 100 & 3.72 & 0.34 & 80.4 & 5.3 & 13.7 & 1.7 &  174.1 & 171.8 \\
Ground-Auto & +X & 208 & 100 & 3.95 & 0.52 & 82.2 & 7.7 & 14.1 & 2.6 &   &  \\
 & -X &  &  & 3.55 & 0.44 & 79.3 & 7.4 & 13.4 & 2.5 &  174.1 & 170.6 \\
Ground-Auto & +Y & 208 & 100 & 3.80 & 0.44 & 77.7 & 7.1 & 14.5 & 2.3 &   &  \\
 & -Y &  &  & 3.58 & 0.56 & 82.6 & 9.0 & 12.3 & 2.6 &  174.5 & 171.0 \\
Ground-Auto  & \multicolumn{11}{|l|}{The Z-hemisphere model failed to produce values for the -Z parameters} \\
\hline
HitL 4.14g & All-sky &  212 & 100 & 3.71 & 0.24 & 90.0 & 3.5 & 14.3 & 2.5 &  241.3 & 238.9 \\
HitL 4.14g & +X & 212 & 100 & 3.76 & 0.32 & 90.7 & 4.8 & 13.8 & 3.6 &   &  \\
 & -X &  &  & 3.68 & 0.33 & 89.7 & 5.0 & 15.4 & 3.7 &  241.3 & 237.4 \\
HitL 4.14g & +Y & 212 & 100 & 3.89 & 0.33 & 89.3 & 5.2 & 13.9 & 3.6 &   &  \\
 & -Y &  &  & 3.53 & 0.30 & 90.5 & 4.5 & 14.9 & 3.7 &  241.5 & 237.5 \\
HitL 4.14g & +Z & 212 & 100 & 3.43 & 0.27 & 88.3 & 4.7 & 12.7 & 2.4 &   &  \\
 & -Z &  &  & 4.36 & 0.52 & 90.4 & 6.0 & 16.2 & 8.2 &  242.5 & 238.5 \\
\hline
HitL 4.13 & All-sky & 211 & 100 & 3.57  & 0.32 & 79.8 & 5.3 & 12.7 & 1.5 & 213.5 & 211.1  \\
HitL 4.13 & +X & 211 & 100 & 3.48 & 0.39 & 83.1 & 6.3 & 12.9 & 2.2 &   &  \\
 & -X &  &  & 3.64 & 0.52 & 75.9 & 8.8 & 12.3 & 2.2 &  213.8 & 210.1 \\
HitL 4.13 & +Y & 211 & 100 & 3.63 & 0.48 & 77.1 & 7.6 & 13.1 & 2.0 &   &  \\
 & -Y &  &  & 3.50 & 0.42 & 81.9 & 7.8 & 11.8 & 2.4 &  213.8 & 210.1 \\
HitL 4.13 & +Z & 211 & 100 & 3.40 & 0.37 & 79.6 & 6.4 & 11.4 & 1.5 &   &  \\
 & -Z &  &  & 3.94 & 0.61 & 76.0 & 9.2 & 15.4 & 3.5 &  214.4 & 210.8 \\
\hline
\end{tabular}
\caption{Hemisphere-dependent core-plus-tail fits to systematic uncertainties on GBM localizations\label{tab:corept_hemi}}}.
\end{table}
\end{appendix}


\begin{thebibliography}{31}
\expandafter\ifx\csname natexlab\endcsname\relax\def\natexlab#1{#1}\fi

\bibitem[{Aasi {et~al.}(2013)}]{abadie2013}
Aasi, J., {et~al.} 2013, arXiv:1304.0670

\bibitem[{Abadie {et~al.}(2010)}]{abadie10}
Abadie, J., {et~al.} 2010, Classical \& QG, 27, 17

\bibitem[{Abadie {et~al.}(2012)}]{abadie12}
---. 2012, Astrophys. J., 760, 12

\bibitem[{Abdo {et~al.}(2009)}]{apj090902b}
Abdo, A.~A., {et~al.} 2009, Astrophys. J. Lett., 706, L138

\bibitem[{Ackermann {et~al.}(2013)}]{lat_catalog}
Ackermann, M., {et~al.} 2013, Astrophys. J. Supp., 209, 11

\bibitem[{Atwood {et~al.}(2009)}]{atwood09}
Atwood, W.~S., {et~al.} 2009, Astrophys. J., 697, 1071

\bibitem[{Band {et~al.}(1993)}]{band}
Band, D.~L., {et~al.} 1993, Astrophys. J., 413, 281

\bibitem[{Bissaldi {et~al.}(2009)}]{bissaldi2009}
Bissaldi, E., {et~al.} 2009, Exp. Astron., 24, 47

\bibitem[{Briggs {et~al.}(1999)}]{briggs99}
Briggs, M.~S., {et~al.} 1999, Astrophys. J. Supp., 503, 122

\bibitem[{Briggs {et~al.}(2009)}]{briggs2008}
---. 2009, AIP Conf.Proc. 1133, ed. Meegan, Gehrels, \& Kouveliotou, 40

\bibitem[{Bromberg {et~al.}(2013)}]{bromberg13}
Bromberg, O., {et~al.} 2013, Astrophys. J., 764, 179

\bibitem[{Connaughton {et~al.}(2013)}]{vc_multimessenger}
Connaughton, V., {et~al.} 2013, EAS Publications Series, 61, 657

\bibitem[{de~Ugarte~Postigo {et~al.}(2013)}]{gcn15187}
de~Ugarte~Postigo, A., {et~al.} 2013, GCN Circular 15187

\bibitem[{Goldstein {et~al.}(2012)}]{goldstein_1grb}
Goldstein, A.~M., {et~al.} 2012, Astrophys. J. Supp., 199, 19

\bibitem[{Graziani \& Lamb(1996)}]{carlo96}
Graziani, C., \& Lamb, D.~Q. 1996, AIP Conf.Proc. 384, ed. Kouveliotou, Briggs,
  \& Fishman, 382

\bibitem[{Gruber {et~al.}(2014)}]{dg_2grb}
Gruber, D., {et~al.} 2014, Astrophys. J. Supp., 211, 12

\bibitem[{Hurley {et~al.}(2013)}]{ipn_catalog}
Hurley, K., {et~al.} 2013, Astrophys. J. Supp., 207, 39

\bibitem[{Kippen {et~al.}(2007)}]{rmk2007}
Kippen, R.~M., {et~al.} 2007, AIP Conf.Proc. 921, 590

\bibitem[{Leloudas {et~al.}(2013)}]{gcn14983}
Leloudas, G., {et~al.} 2013, GCN Circular 14983

\bibitem[{Levan {et~al.}(2013)}]{gcn14455}
Levan, A.~J., {et~al.} 2013, GCN Circular 14455

\bibitem[{Loredo(1990)}]{loredo}
Loredo, T.~J. 1990, in Maximum Entropy and Bayesian Methods, ed. P.~F.
  Foug\`{e}re (Kluwer Academic, Dordrecht), 81

\bibitem[{Meegan {et~al.}(2009)}]{meegan09}
Meegan, C., {et~al.} 2009, Astrophys. J., 702, 791

\bibitem[{Paciesas {et~al.}(2012)}]{paciesas_1grb}
Paciesas, W.~S., {et~al.} 2012, Astrophys. J. Supp., 199, 18

\bibitem[{Pal'shin {et~al.}(2013)}]{valentin}
Pal'shin, V.~D., {et~al.} 2013, Astrophys. J. Supp., 207, 38

\bibitem[{Pandey {et~al.}(2010)}]{pandey090902b}
Pandey, S., {et~al.} 2010, Astrophys. J., 714, 799

\bibitem[{Pendleton {et~al.}(1999)}]{locburst}
Pendleton, G.~N., {et~al.} 1999, Astrophys. J., 512, 362

\bibitem[{Ram\'irez {et~al.}(2013)}]{gcn14685}
Ram\'irez, R.~S., {et~al.} 2013, GCN Circular 14685

\bibitem[{Singer {et~al.}(2013)}]{singer2013}
Singer, L., {et~al.} 2013, Astrophys. J. Lett., 776, 34

\bibitem[{Sivia(1996)}]{sivia}
Sivia, D.~S. 1996, Data Analysis: A Bayesian Tutorial (Oxford, England: Oxford
  Univ. Press)

\bibitem[{Soderberg {et~al.}(2010)}]{soderberg2010}
Soderberg, A.~M., {et~al.} 2010, Nature, 463, 513

\bibitem[{von Kienlin {et~al.}(2014)}]{azk_2grb}
von Kienlin, A., {et~al.} 2014, Astrophys. J. Supp., 211, 13

\end{thebibliography}
\end{document}